\documentclass[twocolumn,onecolappendix]{aastex6}

\usepackage{epsf}
\usepackage{graphicx}
\usepackage{color}
\usepackage{txfonts}
\usepackage{tabularx}
\usepackage{here}
\usepackage{float}
\usepackage{threeparttable}
\usepackage{tabu}
\usepackage{hyperref}
\usepackage{dcolumn}
\usepackage[T1]{fontenc}
\usepackage{aecompl}

\newcolumntype{C}{>{\centering\arraybackslash}X}
\newcolumntype{L}{>{\raggedright\arraybackslash}X}
\newcolumntype{R}{>{\raggedleft\arraybackslash}X}

\newcommand{\oi}{[O\,{\sc i}]}
\newcommand{\oii}{[O\,{\sc ii}]}
\newcommand{\oiii}{[O\,{\sc iii}]}

\newcommand{\Ni}{[N\,{\sc i}]}
\newcommand{\nii}{[N\,{\sc ii}]}

\newcommand{\sii}{[S\,{\sc ii}]}
\newcommand{\siii}{[S\,{\sc iii}]}
\newcommand{\siv}{[S\,{\sc iv}]}

\newcommand{\hei}{He\,{\sc i}}

\newcommand{\neii}{[Ne\,{\sc ii}]}
\newcommand{\neiii}{[Ne\,{\sc iii}]}

\newcommand{\arii}{[Ar\,{\sc ii}]}
\newcommand{\ariii}{[Ar\,{\sc iii}]}
\newcommand{\ariv}{[Ar\,{\sc iv}]}

\newcommand{\ci}{[C\,{\sc i}]}
\newcommand{\clii}{[Cl\,{\sc ii}]}
\newcommand{\cliii}{[Cl\,{\sc iii}]}

\newcommand{\ha}{H$\alpha$}
\newcommand{\hb}{H$\beta$}

\newcommand{\hi}{H\,{\sc i}}

\newcommand{\feii}{[Fe\,{\sc ii}]}
\newcommand{\feiii}{[Fe\,{\sc iii}]}

\newcommand{\kms}{km s$^{-1}$}

\newcommand{\te}{$T_{\rm e}$}
\newcommand{\Ne}{$n_{\rm e}$}

\begin{document}

\title{
Physical properties of the very young PN Hen3-1357 (Stingray Nebula)
based on multiwavelength observations
}

\shorttitle{A Multiwavelength Study of the Stingray Nebula Hen3-1357}
\shortauthors{Otsuka et al.}

\hypersetup{linkcolor=red,citecolor=blue,filecolor=cyan,urlcolor=magenta}

\email{E-mail: otsuka@asiaa.sinica.edu.tw (MO)}

\author{Masaaki Otsuka\altaffilmark{1}} 
\altaffiltext{1}{Institute of Astronomy and Astrophysics
Academia Sinica, 11F of
Astronomy-Mathematics Building, AS/NTU. No.1, Sec. 4,
Roosevelt Rd, Taipei 10617, Taiwan, R.O.C.}

\author{M.~Parthasarathy\altaffilmark{2}} 
\altaffiltext{2}{Indian Institute of Astrophysics, II B
lock Koramangala, Bangalore 560034, Karnataka, India}

\author{A.~Tajitsu\altaffilmark{3}}
\altaffiltext{3}{Subaru Telescope, National Astronomical
Observatory of Japan, 650 N Aohoku Place, Hilo, HI 96720, U.S.A.}

\author{S.~Hubrig\altaffilmark{4}}
\altaffiltext{4}{Leibniz-Institut fuer Astrophysik Potsdam (AIP) ]
An der Sternwarte 12, 14482 Potsdam, Germany}

 \begin{abstract}
  We have carried out a detailed analysis of the interesting and important
  very young planetary nebula (PN) Hen3-1357 (Stingray Nebula) based on
  a unique dataset of optical to far-IR spectra and photometric
  images. We calculated the abundances of nine
  elements using collisionally excited lines (CELs) and recombination lines
  (RLs). The RL C/O ratio indicates that this PN is O-rich, which is
  also supported by the detection of the broad 9/18\,$\mu$m bands from
  amorphous silicate grain. The observed elemental abundances
  can be explained by asymptotic giant branch (AGB) nucleosynthesis
  models for initially 1-1.5\,$M_{\odot}$ stars with $Z$
  = 0.008. The Ne overabundance might be due to the enhancement of
  $^{22}$Ne isotope in the He-rich intershell.
  By using the spectrum of the central star synthesized by Tlusty
  as the ionization/heating source of the PN, we constructed the
  self-consistent photoionization model with Cloudy to the
  observed quantities, and we derived the gas and dust masses,
  dust-to-gas mass ratio, and core-mass of the central star.
  About 80\,\% of the total dust mass is from warm-cold dust component beyond
  ionization front. Comparison with other Galactic PNe indicates
  that Hen3-1357 is
  an ordinary amorphous silicate rich and O-rich gas PN.
  Among other studied PNe, IC4846 shows many similarities in properties
  of the PN to Hen3-1357, although their post-AGB evolution is quite
  different from each other. Further monitoring observations and
  comparisons with other PNe such as IC4846
  are necessary to understand the evolution of Hen3-1357. 
 \end{abstract}

 \section{Introduction}

 Planetary nebula (PN) is the next evolutionary stage
 of asymptotic branch (AGB) stars. PNe consist of a dusty nebula and
 a hot central star evolving toward a white dwarf. So far, over 1000
 PNe in the Galaxy have been identified
 \citep[e.g.,][]{Frew:2008aa}. Among PNe, Hen3-1357 \citep[SAO244567,
 V839 Ara, PN G331.3-12.1, Stingray Nebula,][]{Bobrowsky:1998aa}
 recently attracts lot of attention and has been studied
 actively since the first classification as a post-AGB star done
 by \citet{1989A&A...225..521P}.

 \cite{Parthasarathy:1993aa,1995A&A...300L..25P}
 discovered that Hen3-1357 has a
 young nebula and is going on post-AGB evolution;
 the UV spectrum in 1988 shows the P-Cygni profiles
 of the N\,{\sc v}\,1239/43\,{\AA} and C\,{\sc iv}\,1548/50\,{\AA} lines
 detected in the spectra taken by the International Ultraviolet Explore
 (\emph{IUE}) and the optical spectra in 1990 and 1992 show many nebular
 emission lines. Hen3-1357 is the first object evolving from a B1 type
 post-AGB supergiant into a PN within the extremely short time scale.

 Using a distance of 5.6\,kpc based on an extinction estimate from
 $UBV$ photometry by \citet{Kozok:1985aa},
 \cite{Parthasarathy:1993aa} estimated the luminosity of the central
 star to be 3000\,$L_{\sun}$. \cite{1995A&A...300L..25P} found that the
 effective temperature ($T_{\rm eff}$) of the central star has increased
 from 37\,500\,K to 47\,500\,K during the same period.
 Later, \citet{Parthasarathy:1997aa} estimated $T_{\rm eff}$ =
 50\,000\,K in 1995. A core-mass versus luminosity relation suggests the
 core-mass of 0.55\,$M_{\sun}$. While, the luminosity had faded
 by a factor of three
 in the UV wavelength from 1988 to 1996
 \citep{Parthasarathy:2006aa}. Increasing $T_{\rm eff}$ as fading UV
 flux indicates dropping luminosity, turning out that Hen3-1357 is
 rapidly evolving toward a white dwarf.

 However, it is difficult to explain its evolution and evolutionary time
 scale. \citet{Parthasarathy:1993aa} estimated a
 kinematical age to be $\sim$2700 years by adopting the distance of
 5.6\,kpc, the (bright rim) radius of 0.8{\arcsec} measured using the
 Hubble Space Telescope (\emph{HST}) image \citep{Bobrowsky:1994aa},
 and an expansion velocity of 8\,{\kms} \citep{Parthasarathy:1993aa}.
 According to the H-burning post-AGB evolution for initially
 1.5\,$M_{\sun}$ stars with metallicity \mbox{$Z$ = 0.016} by
 \citet{Vassiliadis:1994aa}, such stars would take
 over 10$^{4}$ years to evolve into the white dwarf cooling track.
 The discrepancy between the observationally estimated and the model
 predicted time scale suggests that Hen3-1357 might have experienced
 an extraordinary post-AGB evolution.

 \cite{Reindl:2014aa} demonstrated that Hen3-1357 has steadily increased
 its $T_{\rm eff}$ from 38\,000\,K in 1988 to a peak value of 60\,000\,K
 in 2002 and cooled again to 55\,000\,K in 2006 based on the stellar UV
 spectra. They proposed late He-flash evolution to explain this rapid
 $T_{\rm eff}$ increment. \cite{Reindl:2017aa} found that $T_{\rm eff}$
 further cooled down, 50\,000\,K in 2015 using the newly obtained the
 \emph{HST} UV spectra of the central star.
 Such a $T_{\rm eff}$ variation is found by
 \citet{Arkhipova:2013aa}, who estimated $T_{\rm eff}$ = 57\,000\,K in
 1990, 55\,000\,K in 1992, and 41\,000\,K in 2011 using the
 {\oiii}\,5007\,{\AA} line intensities relative to the {\hb}. Through a
 comparison with a theoretically calculated late thermal pulse (LTP)
 evolutionary path, \cite{Reindl:2017aa} concluded that Hen3-1357 might have
 experienced a LTP. As \cite{Reindl:2017aa}
 mentioned, however, we should retain that any theoretical LTP models
 cannot yet fully reproduce the observed parameters of the central star
 of Hen3-1357.

 Despite many efforts, the puzzling evolution of Hen3-1357 remains
 a fatal and challenging problem. For
 understanding Hen3-1357, properties of the nebula are crucial because
 the evolutionary history of the progenitor star has been
 imprinted in the nebula, too. Utilizing nebular emission lines, one can
 easily derive elemental abundances such as C/N/O/Ne, which are
 essential key elements to prove AGB nucleosynthesis. The C/O ratio and
 the dust features seen in mid-IR spectra would suggest how much mass of
 the progenitor has gone into the formation of the nebula. It is of
 interest to investigate conditions of gas and dust and derive
 their masses in terms of material recycling in the Galaxy. Thus, nebula
 analysis is complementary for stellar analysis, and properties of the
 nebula can be the basis for understanding both the PN and its central
 star.

 From these reasons, we investigated properties of the nebula
 based on a unique dataset from UV to far-IR wavelengths
 (0.35-140\,{\micron}). We organize this paper as follows. In \S 2,
 we describe our optical high-dispersion spectroscopy using the
 Fiber-fed Extended Range Optical Spectrograph
 \citep[FEROS;][]{Kaufer:1999aa} attached to the MPG ESO 2.2-m telescope
 and the archival mid-IR and far-IR data taken by the \emph{AKARI} and
 \emph{Spitzer} infrared space telescopes. In \S 3, we describe
 nebular abundance analysis. We first report the C/O and N/O ratios
 using the recombination lines of these elements in this PN. We
 compare the observed abundances with the AGB nucleosynthesis models
 to investigate the initial mass of the progenitor star. In \S 4,
 we construct the spectral energy distribution (SED) model using
 photoionization code Cloudy \citep[][version C13.03]{Ferland:2013aa} to
 investigate physical conditions of the nebula and the central star of
 PN (CSPN). We measure broadband magnitudes of the CSPN from the FEROS
 spectrum. We have a brief discussion on the CSPN's SED.
 In \S 5, we compare the observed elemental
 abundances and dust features with those of other PNe in order to
 verify Hen3-1357 as a PN. In \S 6, we summarize our work.

\section{Observations and Data reduction}

  \begin{table}[t!]
   \caption{Observation log for Hen3-1357.  \label{T:obslog}} 
\tablewidth{\columnwidth}
\footnotesize
 \renewcommand{\arraystretch}{0.95}
\tablecolumns{2}
\centering
\begin{tabular}{@{}lc@{}}
\hline\hline
 Telescope/Instrument&Obs-Date\\
                    & (YYYY-MM-DD)\\
\hline
\emph{Spitzer}/IRS    &2005-03-20\\
MPG ESO 2.2-m/FEROS         &2006-04-16\\
\emph{AKARI}/IRC and FIS  &2006-12-31\\
\emph{Spitzer}/IRAC       &2009-04-22\\
 \hline
\end{tabular}
  \end{table}

  We describe the photometric and spectroscopic dataset
  taken by \emph{Spitzer}, \emph{AKARI}, and our FEROS observations. The
  observation log is summarized in Table~\ref{T:obslog}.
  The \emph{AKARI} data were obtained in 2006 May 6 - 2007 Aug 28,
  the middle date is around 2006 Dec 31.

\subsection{Spitzer and AKARI photometry\label{S:IRAC}}

\begin{table}
\caption{Near- to far-IR band flux densities of Hen3-1357. \label{T:IRAC}}
\tablewidth{\columnwidth}
\footnotesize
 \renewcommand{\arraystretch}{0.95}
\tablecolumns{4}
\centering
   \begin{tabular}{@{}r@{\hspace{4pt}}c@{\hspace{1pt}}
    D{p}{ \pm }{-1}@{\hspace{1pt}}D{p}{ \pm }{-1}@{}}
\hline\hline
    \multicolumn{1}{c}{$\lambda_{\rm
    c}$}&Tele/Instr/Band&\multicolumn{1}{c}{$F_
    {\nu}$} &\multicolumn{1}{c}{$F_{\lambda}$}\\
    \multicolumn{1}{c}{({\micron})} &
	&\multicolumn{1}{c}{(mJy)}
	    &\multicolumn{1}{c}{(erg\,s$^{-1}$\,cm$^{-2}$\,{\micron}$^{-1}$)}\\
    \hline
    3.6 & \emph{Spitzer}/IRAC/Band1 & 1.09(+1)~p~5.22(-1)
	    & 2.58(-12)~p~1.24(-13) \\ 
    4.5 & \emph{Spitzer}/IRAC/Band2 & 1.61(+1)~p~5.02(-2)
	    & 2.38(-12)~p~7.42(-15) \\ 
    5.8 & \emph{Spitzer}/IRAC/Band3 & 1.08(+1)~p~1.68(-1)
	    & 9.87(-13)~p~1.53(-14) \\ 
    8.0 & \emph{Spitzer}/IRAC/Band4 & 3.97(+1)~p~8.20(-1)
	    & 1.89(-12)~p~3.91(-14) \\ 
    9.0 & \emph{AKARI}/IRC/S9W & 8.87(+1)~p~8.62(0) & 3.13(-12)~p~3.04(-13) \\ 
    65.0 & \emph{AKARI}/FIS/N60 & 2.25(+3)~p~3.52(+2)
	    & 1.60(-12)~p~2.50(-13) \\ 
    90.0 & \emph{AKARI}/FIS/WIDE-S& 1.88(+3)~p~5.06(+1)
	    & 6.98(-13)~p~1.87(-14) \\ 
    140.0 & \emph{AKARI}/FIS/WIDE-L & 3.77(+2)~p~2.75(+2)
	    & 5.77(-14)~p~4.21(-14)\\
\hline
 \end{tabular}
\end{table}

 We measured the mid-IR flux densities for Bands 1-4 of the
 \emph{Spitzer}/Infrared Array Camera \citep[IRAC;][]{Fazio:2004aa},
 where the central wavelength ($\lambda_{\rm c}$) is 3.6, 4.5, 5.8, and
 8.0\,{\micron}, respectively.
 We reduced the basic calibrated data (BCD, program-ID: 50116,
 obs AORKEY: 25445376, PI: G.~Fazio) using mosaicking and point-source
 extraction software ({\sc MOPEX})
 \footnote{\url{http://irsa.ipac.caltech.edu/data/SPITZER/docs/dataanalysistools/tools/mopex/}
 }
 provided by Spitzer Science Center (SSC) to create a mosaic image for
 each band. We subtracted artificial features seen in the images as
 possible as we can. After  we had subtracted out surrounding stars by
 point-spread function fittings using the {\sc Digiphot} photometry
 package in {\sc IRAF} v.2.16\footnote{{\sc IRAF} is distributed by the
 National Optical Astronomy Observatories, operated by the Association
 of Universities for Research in Astronomy (AURA), Inc., under a
 cooperative agreement with the National Science Foundation.
 \url{http://iraf.noao.edu}
 }, we
 performed aperture photometry. The results are summarized in
 Table~\ref{T:IRAC}.

 To trace amorphous silicate feature seen in the \emph{Spitzer}/IRS
 spectrum, we used the \emph{AKARI} Infrared Camera
 \citep[IRC;][]{Onaka:2007aa} S9W (\mbox{$\lambda_{c}$ = 9\,{\micron}})
 and L18W (\mbox{$\lambda_{c}$ = 18\,{\micron}}). We used the
 \emph{AKARI} Far-Infrared Surveyor  \citep[FIS;][]{Kawada:2007aa} data
 as vital constraints to the warm-cold dust continuum in the SED
 modeling. For this end, we utilized the photometry measurements by
 \citet{Yamamura:2010aa} for the IRC two bands
 and FIS Bright Source Catalogue Ver.2 for the FIS N60, WIDE-S, and
 WIDE-L bands at $\lambda_{c}$ = 65, 90, and 140\,{\micron},
 respectively. These data were taken by the
 \emph{AKARI} all-sky survey. We list these flux densities
 in Table~\ref{T:IRAC}, where $A$($-B$) means $A\times10^{-B}$ and hereafter.

  \subsection{MPG ESO 2.2-m FEROS spectroscopy
\label{S-feros}
  }

\begin{figure*}[t!]
\centering
\includegraphics[width=0.7\textwidth]{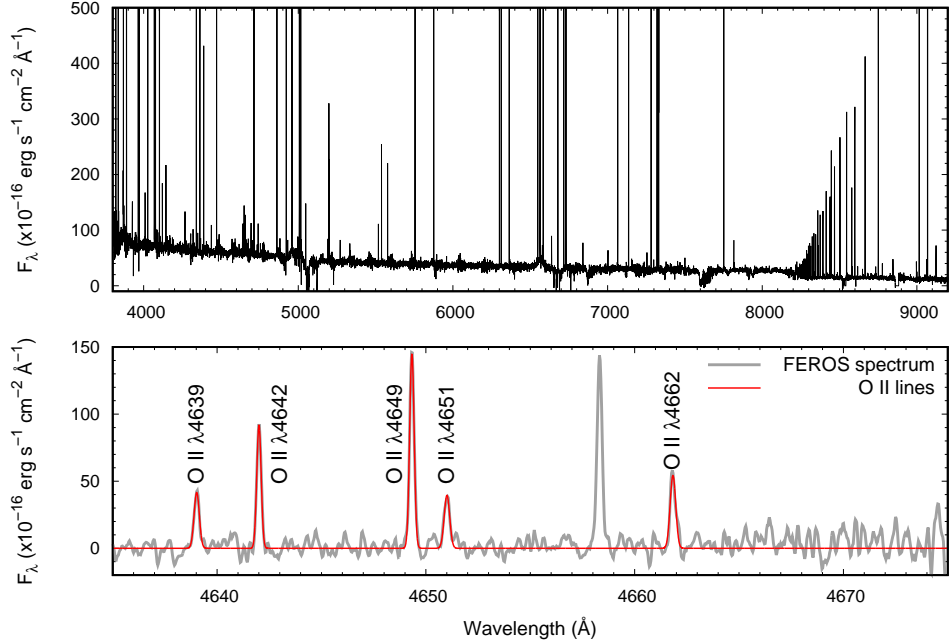}
\caption{
 ({\it upper panel}) The FEROS spectrum of Hen3-1357 in the range between
 3800\,{\AA} and 9200\,{\AA}. ({\it lower panel}) The
 FEROS spectrum in 4635-4675\,{\AA} (grey line)
 and the Gaussian fitting results for the O\,{\sc ii} lines in this
 wavelength range (red line). The local continuum was subtracted out.
\label{S:feros}
}
\end{figure*}

 We secured the optical high-dispersion spectrum
 (3500-9200\,{\AA}) using the FEROS
 attached to the MPG ESO 2.2-m Telescope, La Silla, Chile
 (Prop.ID: 77.D-0478A, PI: M.~Parthasarathy).

 The weather condition was stable and clear throughout the night, and
 the seeing was 0.8-1.17{\arcsec} (average: 0.97{\arcsec})
 measured from the differential image motion monitor. FEROS's fibers use
 2.0{\arcsec} apertures and provide simultaneously
 the object and sky spectra. The detector is the EEV CCD
 chip with \mbox{2048$\times$4096} pixels of
 \mbox{15$\times$15\,{\micron}} square. We selected a \mbox{1$\times$1}
 on-chip binning and
 low gain mode\footnote{We measured the gain = 4.99\,$e^{-}$\,ADU$^{-1}$
 and readout-noise = 8.31\,$e^{-}$
 using the {\sc IRAF} task {\sc Findgain}}. The atmospheric
 dispersion corrector (ADC) was not used during the observation.
 The exposure time was a single 2100 sec at airmass of 1.297-1.380.
 For the flux calibration and blaze function correction, we observed the
 standard star HR\,3454 \citep{Hamuy:1992aa,Hamuy:1994aa} at airmass
 $\sim$1.2.
 Since we did not use the ADC, a color-dependent displacement
 of the source from differential atmospheric refraction (DAR) might be
 present. However, we took Stingray nebula and HR\,3454 at similar
 airmass. Therefore, we believe that DAR effect on the inferred
 extinction coefficients, the derived electron temperatures, and
 therefore on the derived ionic and elemental abundances would be
 largely reduced. 
 We reduced the data with the echelle spectra reduction package {\sc
 Echelle} in {\sc IRAF} by a standard reduction manner including bias
 subtraction, removing scattered light,
 detector sensitivity correction, removing cosmic-ray hits, airmass
 extinction correction, flux density calibration, and all echelle
 order connection. Using the sky spectrum, we subtracted the sky-lines
 from the Hen3-1357 spectrum. The average resolving power
 ($\lambda$/$\Delta\lambda$) is 44\,950,
 which was measured from the average full width at half maximum (FWHM)
 of over 300 Th-Ar comparison lines obtained for the wavelength
 calibrations. The signal-to-noise ratios per pixel were $\sim$2-12 for
 the continuum.

 The resultant FEROS spectrum is presented in Fig.~\ref{S:feros};
 the detected recombination lines (RLs) of O\,{\sc ii} are
 shown in the
 lower panel. As far as we know, the N and O RLs such as N\,{\sc ii} and
 O\,{\sc ii} are detected in this PN for the first time.

\subsection{Spitzer/IRS spectrum}

\begin{figure*}
\centering
\includegraphics[width=0.70\textwidth,clip]{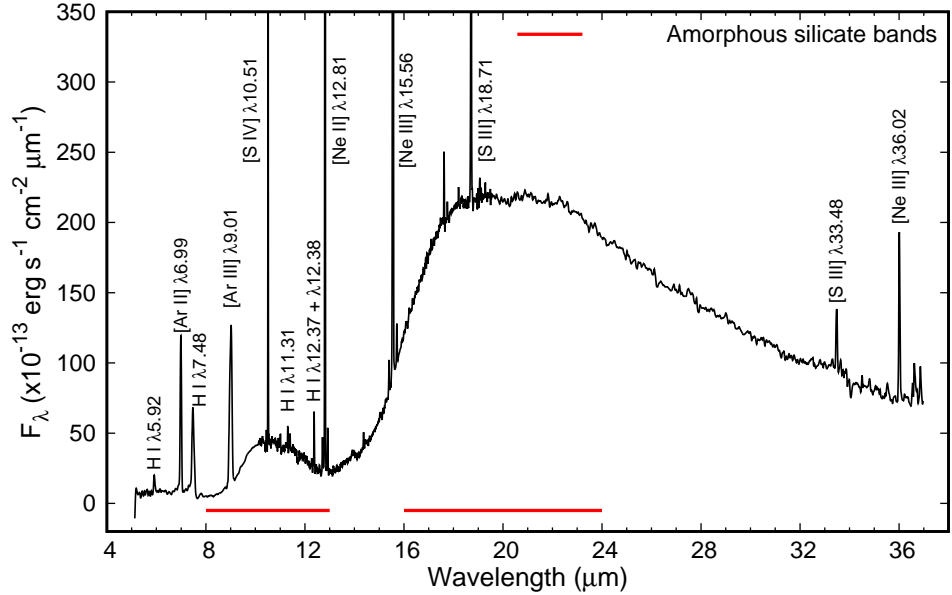}
\caption{
The \emph{Spitzer}/IRS spectrum of Hen3-1357. The identified atomic lines
 and amorphous silicate features are denoted. \label{F:spt}}
\end{figure*}

To investigate dust features and perform plasma diagnostics
using fine-structure lines, we analyzed the mid-IR spectra taken by the 
\emph{Spitzer}/Infrared Spectrograph \citep[IRS;][]{Houck:2004aa} with
the Short-Low (SL, 5.2-14.5\,{\micron}, 
the slit dimension: $\sim$3.6{\arcsec} $\times$ 57{\arcsec}), 
Short-High (SH, 9.9-19.6\,{\micron}, 4.7{\arcsec} $\times$
11.3{\arcsec}), and Long-High modules (LH, 18.7-37.2\,{\micron},
11.1{\arcsec} $\times$ 22.3{\arcsec}).

We processed the BCD (program-ID: 3633, obs AORKEY: 11312640,
PI: B.~Matthew) using the data reduction packages 
{\sc SMART} v.8.2.9 \citep{Higdon:2004aa} and {\sc IRSCLEAN}
v.2.1.1\footnote{
\url{http://irsa.ipac.caltech.edu/data/SPITZER/docs/dataanalysistools/tools/irsclean/}
} provided
by the SSC. We scaled the flux density 
of the reduced LH-spectrum to match with that of the reduced SH-spectrum in 
the overlapping wavelength, and we obtained the single 
9.9-37.2\,{\micron} spectrum. Then, by the similar way, we combined this
high-dispersion spectrum and the SL 5.2-14.5\,{\micron} 
spectrum into the single 5.2-37.2\,{\micron} spectrum.

We present the resultant spectrum in Fig.~\ref{F:spt}. The intensity 
peak positions of the identified atomic lines are marked by the vertical
lines. We detected Ne, S, and Ar fine-structure lines. 
The spectrum clearly shows two broad features (indicated by the horizontal 
red lines) attributed  to amorphous silicate grains; the features centered 
at 9\,{\micron} and 18\,{\micron} are due to the Si-O stretching
mode and the O-Si-O bending mode,
respectively. \citet{Perea-Calderon:2009aa} reported that this
PN is an O-rich dust object. We did not identify any carbon-based dust 
grains and molecules in the \emph{Spitzer}/IRS spectrum.
Thus, we can conclude that Hen3-1357 has an O-rich dust nebula.

\section{Results}

  \subsection{Scaling the flux density of the \emph{Spitzer}/IRS spectrum
  \label{S:Spitzer}}

We performed a correction to recover the loss of light from
Hen3-1357 by the slit.

First, using the \emph{AKARI}/IRC 9.0\,{\micron}
band photometry listed in Table~\ref{T:IRAC}, we scaled the flux
density of the spectrum by considering the \emph{AKARI}/IRC
9.0\,{\micron} filter transmission curve by a constant scaling
factor of 0.951. Next, using this scaled
spectrum and the \emph{Spitzer}/IRAC 8.0\,{\micron} filter transmission curve, 
we measured the \emph{Spitzer}/IRAC 8.0\,{\micron} band flux density. The measured value 
\mbox{1.90(--12)~erg\,s$^{-1}$\,cm$^{-2}$\,{\micron}$^{-1}$} is consistent with that
the IRAC 8.0\,{\micron} photometry result.

\emph{AKARI}/IRC 9.0\,{\micron} and \emph{Spitzer}/IRAC 8.0\,{\micron}
bands include atomic lines of H, Ne, S, and Ar certainly
contributing to these two bands. As noted in
\S \ref{S:2006a2011}, we did not find a significant difference between optical nebular
line intensities relative to the {\hb} measured in 2006 and
in 2011. This means that the ionization and elemental abundances of the
nebula might not be changed in 2006-2011.

Our adopted scaling factor (0.951) indicates that the IR-band flux
decreased by $\sim$5\,\% between 2005 and 2009.
Therefore, we assume that mid-IR
wavelength evolution had not dramatically changed in 2005-2009.

Taking into account these analyses, we scaled the flux density
of the spectrum to match with the \emph{AKARI}/IRC 9.0\,{\micron} 
band flux density.

\subsection{The H$\beta$ flux of the entire nebula \label{S:Hb}}

The {\hb} flux of the entire nebula is necessary for
setting the nebula's hydrogen density structure in our SED modeling 
as well as for calculating the Ne$^{+,2+}$, S$^{2+,3+}$, and 
Ar$^{+,2+}$ to H$^{+}$ number density ratios, and electron density {\Ne}
and temperature {\te} using mid-IR fine-structure lines of these ions.

Since the H\,{\sc i}\,7.46\,{\micron} line 
is in the longer wavelength edge of the SL2 spectrum 
(5.13-7.60\,{\micron}) and also in the shorter edge of the SL1 spectrum (7.46-14.29\,{\micron}),
we did not employ this line for estimating the {\hb} line flux of the 
entire nebula. Therefore, we obtained the {\hb} line flux 
by utilizing the theoretical H\,{\sc i} $I$($n$ = 7-6 and 11-8)/$I$($n$
= 4-2) intensity ratio calculated 
by \citet{Storey:1995aa}, where $n$ is the principal quantum
number. Note that a detected line at 12.37\,{\micron} (see
Fig.~\ref{F:spt}) indeed composes of the H\,{\sc i} 
\mbox{$n$ = 7-6} at 12.37\,{\micron} and \mbox{$n$ = 11-8} at 12.38\,{\micron}. 
From the $I$(12.37\,{\micron} + 12.38\,{\micron})/$I$({\hb}) 
= 1.04(--2) in the case of an {\Ne} = 10$^{4}$\,cm$^{-3}$ 
and a {\te} = 10$^{4}$\,K \citep{Storey:1995aa}, we estimated 
the {\hb} flux of the entire nebula to be 
\mbox{9.83(--12)~$\pm$~7.33(--13)\,erg~s$^{-1}$~cm$^{-2}$}.

\subsection{Flux measurements \label{S:flux}}

\begin{table}[t!]
 \caption{The derived $c$(H$\beta$) ratios. For the interstellar
 reddening correction to the FEROS spectrum, we adopted the average $c$({\hb})
= \mbox{8.27(--2)~$\pm$~3.47(--2)} \label{T:chb}}
\renewcommand{\arraystretch}{0.95}
 \tablecolumns{3}
\tablewidth{\columnwidth}
\footnotesize
\centering
\begin{tabular}{@{}c@{\hspace{5pt}}l@{\hspace{5pt}}c@{\hspace{8pt}}@{}}
\hline\hline
$\lambda_{\rm lab.}$ ({\AA})&Line&$c$({\hb})\\
\hline
3797.9 &B10 &8.25(--2) $\pm$ 6.43(--3) \\
3835.4 &B9 &3.25(--2) $\pm$ 5.17(--3) \\
3970.1 &B7 &7.39(--2) $\pm$ 2.95(--3) \\
4101.7 &B6 &4.38(--1) $\pm$ 3.56(--3) \\
4340.5 &B5 &1.34(--1) $\pm$ 5.16(--3) \\
8545.4 &P15 &2.03(--2) $\pm$ 1.04(--2)\\
8598.4 &P14 &3.84(--2) $\pm$ 2.54(--3)\\
8665.0 &P13 &7.49(--2) $\pm$ 2.48(--3)\\
8750.5 &P12 &6.64(--2) $\pm$ 1.98(--3)\\
9014.9 &P10 &1.18(--2) $\pm$ 2.25(--3)\\ 
 \hline
\end{tabular}
\end{table}

We measured the fluxes of the emission lines by Gaussian fittings. Then, 
we corrected these fluxes using the following formula;

\begin{equation}
 I(\lambda) = F(\lambda)~\cdot~10^{c({\rm H\beta})(1 + f(\lambda))},
\end{equation}
 
\noindent
where $I$($\lambda$) is the de-reddened line flux, $F$($\lambda$)
is the observed line flux, $f$($\lambda$) is the interstellar
extinction function at $\lambda$ computed by the reddening law of
\citet{Cardelli:1989aa} with $R_{V}$ = 3.1, $c$({\hb}) is the
reddening coefficient at {\hb}.

We measured $c$({\hb}) values by comparing the observed ten
Balmer and Paschen line ratios to {\hb} 
with the theoretical ratios of \citet{Storey:1995aa} for a 
{\te} = 10$^{4}$\,K and an {\Ne} = 10$^{4}$\,cm$^{-3}$ under the Case B
assumption. To reduce $c$({\hb}) estimation errors originated from
the {\hi} absorptions in the flux standard star HR\,3454,
we estimated $c$({\hb}) using different line ratios.
The derived $c$({\hb}) values are listed in
Table~\ref{T:chb}. Since the {\ha} line was saturated, we did not calculate a
$c$({\hb}) using the $F$({\ha})/$F$({\hb}) ratio. Finally, we
adopted the average $c$({\hb})
= \mbox{8.27(--2)~$\pm$~3.47(--2)}. The scatter between the
estimated $c$({\hb}) could be due to the {\hi} absorptions' depth of
HR\,3454 measured by \citet{Hamuy:1992aa,Hamuy:1994aa}.
We did not correct interstellar extinction for the \emph{Spitzer}/IRS spectrum because
the extinction is negligibly small in mid-IR wavelength.

For the year 2006, \citet{Reindl:2014aa} reported $E(B-V)$ =
0.11, corresponding to $c$({\hb}) = 0.16.
Although they did not give the uncertainty of $E(B-V)$, we assume
$\delta\,E(B-V)$ = 0.02 from the fact that they measured $E(B-V)$ =
0.14 $\pm$ 0.02 in the year 1997. Thus, their $c$({\hb}) for the year
2006 is estimated to be 0.16 $\pm$ 0.03, which is consistent with ours.

In appendix Table~\ref{T:feros}, we list 180 nebular lines detected in the FEROS
spectrum. Since the {\oiii}\,5007\,{\AA} and {\ha} lines were
saturated, we do not list their fluxes.  
We calculated the average heliocentric radial velocity 12.30\,km
s$^{-1}$ and local standard of rest (LSR)
radial velocity 12.29\,km s$^{-1}$ using all the identified lines in
the FEROS spectrum (1-$\sigma$ uncertainty is 0.25\,km s$^{-1}$). Our
heliocentric radial velocity
is in good agreement with \citet[][\mbox{12.6~$\pm$~1.7\,{\kms}}]{Arkhipova:2013aa}.

In appendix Table~\ref{T:spitzer}, we listed the fluxes of the identified 
14 atomic gas emission-lines detected in 
the {\it flux density scaled} \emph{Spitzer}/IRS spectrum, where the
fluxes are normalized with respect to the {\hb} flux of the entire nebula.

\subsection{Comparison of line fluxes between 2006 and 2011 \label{S:2006a2011}}

We investigated the possibility of temporal variations of the emission
line intensities by comparing our measurements with those of
\citet{Arkhipova:2013aa}, who obtained
the 3500-7200\,{\AA} low-resolution spectrum (FWHM = 4.5\,{\AA}) on 2011
June at the South African Astronomical Observatory (SAAO).
In appendix
Table~\ref{T:comp_obs}. We list
their measured line-intensities overlapped with ours. In 
2006-2011, the nebular line fluxes did not significantly change. Indeed,
the $I$($\lambda$)s in 2006 are very similar to those in 2011
($I$(2011)/$I$(2006) = \mbox{1.11~$\pm$~0.02}, correlation factor is 0.995).
Thus, the ionization and elemental abundances of the
nebula might not be largely changed in 2006-2011.
Variation in the $T_{\rm eff}$ of the central star
by 5000\,K to 10\,000\,K in 5 to 10 years interval might
not immediately change the nebular morphology, parameters and
abundances in the same time period.

\subsection{Plasma-diagnostics}

\begin{figure*}[t!]
\centering
\includegraphics[width=0.7\textwidth,clip]{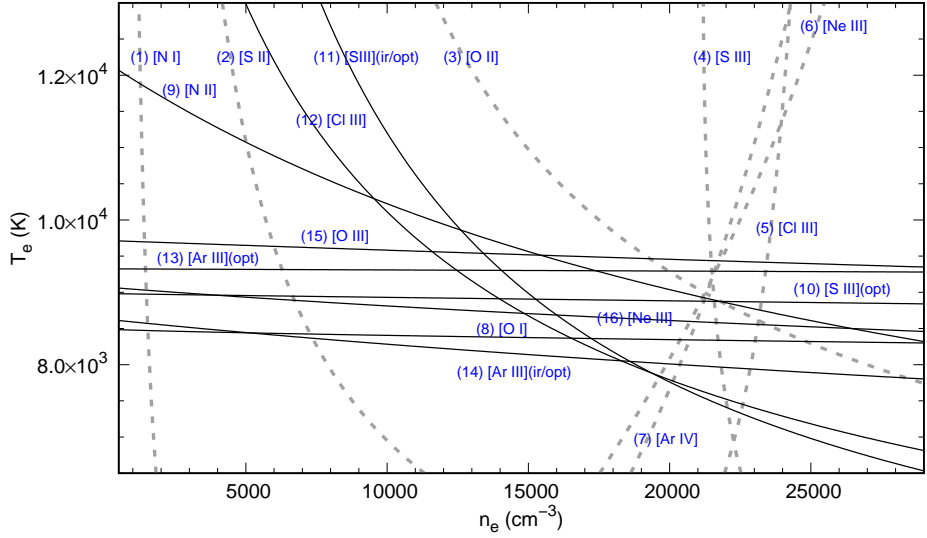}
\caption{
 {\Ne}-{\te} diagram based on CEL diagnostic line ratios.
 The dashed and thick lines with ``(ID)`` indicate the {\Ne} and
 {\te}-diagnostic curves generated by the line ratios listed
 in Table~\ref{T:diagno}, respectively.
\label{F:diagno}
 }
\end{figure*}

\begin{table*}[t!]
 \caption{Summary of plasma-diagnostics.
 As a comparison, the results by
 \citet[][for the year 2011]{Arkhipova:2013aa} are listed in the last
 column. \label{T:diagno}}
\footnotesize
 \centering
 \renewcommand{\arraystretch}{0.95}
\tablecolumns{6}
\tablewidth{\columnwidth}
 \begin{tabular}
  {@{}lllD{p}{\pm}{-1}D{p}{\pm}{-1}D{p}{\pm}{-1}@{}}
\hline\hline
\multicolumn{5}{c}{{\Ne}-derivations (this work for the year
  2006)}&\multicolumn{1}{c}{\citet{Arkhipova:2013aa}}\\
  \cline{1-5}
ID   &Ion&Diagnostic line ratio&\multicolumn{1}{c}{Ratio} &\multicolumn{1}{c}{Result (cm$^{-3}$)}&\multicolumn{1}{c}{(cm$^{-3}$)}\\
\hline
(1)    &{\Ni} &$I$(5197\,{\AA})/$I$(5200\,{\AA}) & 1.59(0)~p~3.16(-2) & 1390~p~90 &$\nodata$\\ 
(2)    &{\sii} &$I$(6716\,{\AA})/$I$(6731\,{\AA}) & 3.23(-1)~p~7.30(-2) & 5710~p~1790&8740~p~7701 \\ 
(3)   &{\oii} &$I$(3726/29{\AA})/$I$(7320/30\,{\AA}) & 4.37(0)~p~1.17(-1) & 17\,520~p~530 &$\nodata$\\ 
(4)   &{\siii} &$I$(18.7\,{\micron})/$I$(33.5\,{\micron}) & 3.94(0)~p~3.32(-1) & 21\,990~p~4840 &$\nodata$\\ 
(5)  &{\cliii} &$I$(5517\,{\AA})/$I$(5537\,{\AA}) & 4.86(-1)~p~1.69(-2) & 23\,970~p~3120 &$\nodata$\\ 
(6)    &{\neiii} &$I$(15.6\,{\micron})/$I$(36.0\,{\micron}) & 1.56(+1)~p~9.67(-1) & 22\,750~p~5850 &$\nodata$\\ 
(7)   &{\ariv} &$I$(4711\,{\AA})/$I$(4740\,{\AA}) & 4.88(-1)~p~4.55(-2) &22\,720~p~4360 &$\nodata$\\ 
            &H\,{\sc i} & Paschen decrement       &&\multicolumn{1}{r}{10\,000 -- 20\,000}&$\nodata$\\ 
\hline
\multicolumn{5}{c}{{\te}-derivations (this work for the year
  2006)}&\multicolumn{1}{c}{\citet{Arkhipova:2013aa}}\\
  \cline{1-5}
ID     &Ion               &Diagnostic line ratio&\multicolumn{1}{c}{Ratio} &\multicolumn{1}{c}{Result (K)}&\multicolumn{1}{c}{(K)}\\
\hline
(8)    &{\oi} &$I$(6300/63\,{\AA})/$I$(5577\,{\AA}) & 9.69(+1)~p~2.57(0) & 8470~p~70&$\nodata$ \\ 
(9)   &{\nii} &$I$(6548/83\,{\AA})/$I$(5755\,{\AA}) & 6.38(+1)~p~1.55(0) & 9280~p~100&11\,066~p~1752 \\ 
(10)  &{\siii} &$I$(9069\,{\AA})/$I$(6312\,{\AA}) & 1.22(+1)~p~6.34(-1) & 8880~p~180 &11\,831~p~2286\\ 
(11)  &{\siii} &$I$(18.7/33.5\,{\micron})/$I$(9069\,{\AA}) & 1.38(0)~p~7.82(-2) & 7430~p~280&$\nodata$ \\ 
(12)  &{\cliii} &$I$(5517/37\,{\AA})/$I$(8434/8501\,{\AA}) & 2.03(+1)~p~4.54(0) & 7490~p~850&$\nodata$ \\ 
(13)  &{\ariii} &$I$(7135/7751\,{\AA})/$I$(5191\,{\AA}) & 2.09(+2)~p~1.14(+1) & 8670~p~150 &$\nodata$\\ 
(14)  &{\ariii} &$I$(9.01\,{\micron})/$I$(7135/7751\,{\AA}) & 9.70(-1)~p~4.27(-2) & 8400~p~310&$\nodata$ \\ 
(15)  &{\oiii} &$I$(4959\,{\AA})/$I$(4363\,{\AA}) & 5.91(+1)~p~7.27(-1) & 9420~p~40 &11\,553~p~1579\\ 
(16) &{\neiii} &$I$(15.6\,{\micron})/$I$(3869/3968\,{\AA}) & 1.92(0)~p~7.34(-2) & 8560~p~70 &$\nodata$\\ 
{\te}(PJ) &&($I_{\lambda}$(8194\,{\AA})-$I_{\lambda}$(8169\,{\AA}))/$I$(P11)  & 2.16(-2)~p~2.53(-3) & 8090~p~1680 &$\nodata$\\ 
  {\te}(He\,{\sc i}) &{\hei}& $I$(7281\,{\AA})/$I$(6678\,{\AA})  & 1.83(-1)~p~7.38(-3) & 8340~p~330&$\nodata$ \\ 
{\te}(He\,{\sc i}) &{\hei}& $I$(7281\,{\AA})/$I$(5876\,{\AA})  & 4.95(-2)~p~1.80(-3) & 7980~p~360&$\nodata$ \\ 
\hline
\end{tabular}
\end{table*}

In forbidden line analysis, we employed the NEBULAR package by
 \citet{Shaw:1995aa}. In recombination line analysis, 
we used private softwares. In both of emission line analyses, we adopted 
effective recombination coefficients, transition 
probabilities, and effective collision strengths listed in 
\citet[][their Table 7]{Otsuka:2010aa}.

We performed plasma diagnostics using collisionally excited lines (CELs)
and RLs. We greatly increase
the results, comparing to \citet{Parthasarathy:1993aa} who obtained one
{\Ne} and two {\te} using the optical spectrum taken in 1992 and
\citet{Arkhipova:2013aa} who deduced one {\Ne} and four {\te} based on
the 3500-7200\,{\AA} spectrum taken in 2011.
In Table~\ref{T:diagno}, we list the diagnostic line ratios to
derive {\Ne} and {\te} and the resultant values.
In Fig.~\ref{F:diagno}, we present the {\Ne}-{\te} diagram
using the diagnostic CEL ratios. ``opt'' indicates the result by 
the optical forbidden line ratio; e.g., {\siii}
 $I$(9069\,{\AA})/$I$(6312\,{\AA}) ratio. ``ir/opt'' means the result using 
the mid-IR fine-structure lines and optical forbidden line; e.g., {\siii}
 $I$(18.7/33.5\,{\micron})/$I$(9067\,{\AA}) ratio.
 We bear in mind that CEL emissivities are in general sensitive to {\Ne}
 and {\te}, accordingly CEL ionic abundances depend on a selection of {\Ne} and {\te}.

First, we calculated {\Ne} using CELs. The {\Ne}-{\te} diagram 
indicates that the average {\Ne} is in a range from 
$\sim$2000\,cm$^{-3}$ in neutral gas regions (by the
{\Ne}({\Ni}) curve, ID(1)) and $\sim$20\,000\,cm$^{-3}$ in highly
ionized gas regions (by the {\Ne}({\ariv}) curve, ID(7)) and the average {\te}
is $\sim$8000-10\,000\,K. We derived all {\Ne} by adopting a constant {\te} = 9\,000\,K.

Next, we calculated {\te}({\oi}) by adopting {\Ne}({\Ni}),
{\te}({\ariii}) by the average {\Ne} = 22\,980\,cm$^{-3}$ between {\Ne}({\siii}) and {\Ne}({\cliii}),
{\te}({\siii}) by {\Ne}({\siii}), {\te}({\cliii}) by {\Ne}({\cliii}),
and both {\te}({\oiii}) and {\te}({\neiii}) by adopting {\Ne}({\neiii}), respectively.

To obtain {\te}({\oii}), {\Ne}({\oii},
and {\te}({\nii}) which are representative {\Ne} and {\te} in lower
ionization regions, we subtracted respective contributions from O$^{2+}$ and 
N$^{2+}$ recombination to the {\oii}\,7320/30\,{\AA} lines and the
{\nii}\,5755\,{\AA} line. We calculated the contributions to these lines, 
\mbox{$I_{\rm R}$({\oii}\,7320/30\,{\AA})} and \mbox{$I_{\rm R}$({\nii}\,5755\,{\AA})}
by using the following equations by \citet{Liu:2000aa}.

\begin{equation}
\frac{I_{\rm R}({\rm [O\,{\sc II}]\,7320/30\,{\AA}})}{I({\rm H\beta})} 
= 9.36\left(\frac{T_{\rm e}}{10^{4}}\right)^{0.44}\frac{n({\rm O^{2+}})}{n({\rm H^{+}})},
\end{equation}

\begin{equation}
\frac{I_{\rm R}({\rm [N\,{\sc II}]\,5755\,{\AA}})}{I({\rm H\beta})} = 
3.19\left(\frac{T_{\rm e}}{10^{4}}\right)^{0.33}\frac{n({\rm N^{2+}})}{n({\rm H^{+}})}.
\end{equation}

\noindent Here \mbox{$n$(O$^{2+}$)/$n$(H$^{+}$)} and 
\mbox{$n$(N$^{2+}$)/$n$(H$^{+}$)} are the number density ratios of
the O$^{2}$ and N$^{+}$ with respect to the H$^{+}$,
respectively.

We adopted the CEL O$^{2+}$ = \mbox{1.87(--4)~$\pm$~1.39(--6)}
(see \S \ref{S:ionic}) in order to obtain the
\mbox{$I_{\rm R}$({\oii}\,7320/30\,{\AA})} = \mbox{0.16~$\pm$~0.01}, where $I$({\hb}) = 100. 
Based on the result that the CEL O$^{2+}$ is
consistent with the RL O$^{2+}$, we assumed that the CEL N$^{2+}$ could be 
very close to the RL N$^{2+}$. Here, we adopted the RL N$^{2+}$ =
\mbox{6.97(--5)~$\pm$~3.17(--5)}
(see \S \ref{S:ionic}) to calculate the \mbox{$I_{\rm R}$({\nii}\,5755\,{\AA})} of \mbox{0.33~$\pm$~0.05}.

The {\oii} \mbox{$I$(3726\,{\AA})/$I$(3729\,{\AA})} ratio is a {\Ne}
indicator and \mbox{$I$(3726/29\,{\AA})/$I$(7320/30\,{\AA})} ratio is
sensitive to both {\te} and {\Ne}. In Hen3-1357,
{\Ne} exceeds the critical density of the
\mbox{{\oii}\,3726/29\,{\AA}} lines, so that the \mbox{$I$(3726\,{\AA})/$I$(3729\,{\AA})} ratio could not give reliable
{\Ne}. Therefore, we used the \mbox{$I$(3726/29\,{\AA})/$I$(7320/30\,{\AA})} 
ratio to derive an {\Ne} required for the N$^{+}$, O$^{+}$,
Cl$^{+}$, Ar$^{+}$, and Fe$^{2+}$ calculations. We obtained
{\Ne}({\oii}) by adopting a constant
{\te} = 9000\,K, and then {\te}({\nii}) using this {\Ne}({\oii}) = 17\,520\,cm$^{-3}$.

We found the discrepancy between two {\te}({\siii}) values (IDs 10 and
11). This might be due to the underestimated {\siii}\,9069\,{\AA}, which
is appeared in the red wavelength edge of the FEROS spectrum.
Because the ionic S$^{2+}$ abundance from this line is $\sim$14\,$\%$
smaller than that from the fine-structure {\siii} lines, which
is insensitive to {\te} (see Table~\ref{T:CEL}).
As we explained in \S,\ref{S-feros}, the
differential atmospheric refraction (DAR) effect
might have affected {\siii}\,9069\,{\AA}, although we cannot exactly estimate
how much light of {\siii}\,9069\,{\AA} line we lost.
DAR effect might affect widely separated diagnostic line intensity
ratios. However, for the S$^{2+}$ abundance estimate,
we adopted the average {\te} between two {\te}({\siii}). Thus,
we reduced the effects by inconsistency between these two
{\te}({\siii}).

Similarly, if we underestimate the
[O\,{\sc ii}]\,7320/30\,{\AA} intensity by $\sim$14\%, which is
an expected value from the above analysis for the S$^{2+}$ abundance,
we obtain {\Ne}({\oii}) = 20\,300 cm$^{-3}$. Then using this
{\Ne}({\oii}), we obtain {\te}({\nii}) = 9010\,K. Under
these {\Ne}({\oii}) and {\te}({\nii}), the N$^{+}$,
O$^{+}$, Cl$^{+}$, and Fe$^{2+}$ abundances\footnote{We
calculated these ionic abundances under the {\Ne}({\oii}) and
{\te}({\nii}). See appendix Table \ref{T:nete}
} would increase by $\sim$12\,\%.
Even if DAR effect is present in our FEROS spectrum, the potential
error of $c$({\hb}), {\te} and {\Ne}, and ionic/elemental abundances
caused by DAR effect would be $\sim$15\,\% or less. Hence, our
conclusion on these physical parameters derived from the CELs and the
RLs does not change.

Finally, we calculated {\te} and {\Ne} using He\,{\sc i} lines and H\,{\sc i} Paschen
series. We calculated {\te}({\hei}) using
{\hei} \mbox{$I$(7281\,{\AA})/$I$(6678\,{\AA})} and
\mbox{$I$(7281\,{\AA})/$I$(5876\,{\AA})} ratios using the recombination 
coefficients in a constant {\Ne} = 10$^{4}$\,cm$^{-3}$ provided 
by \cite{Benjamin:1999aa}. We calculated the Paschen 
jump {\te}(PJ) by using equation (7) of \citet{Fang:2011aa}. The H\,{\sc
i} P11 line is in an echelle order gap. Therefore, we obtained the {\it
expected} $I$(P11) using the observed H\,{\sc i} P12
line and the theoretical \mbox{$I$(P11)/$I$(P12)} ratio of 1.30 in {\Ne}
$\sim$10$^{2}$-10$^{5}$\,cm$^{-3}$ and {\te}
$\sim$5000-15\,000\,K \citep{Storey:1995aa}. Thus, we obtained
{\Ne} $\sim$10\,000-20\,000\,cm$^{-3}$ by comparing the
observed \mbox{$I$(P$n$)/$I$(P10)} ratios ($n$ is from 12 to 42) 
and the theoretical calculations under
the Case B assumption and {\te}(PJ) = 8090\,K by \citet{Storey:1995aa}.

As a comparison, the results by \citet[][for the year 2011]{Arkhipova:2013aa} are listed in the last column.
\citet{Arkhipova:2013aa} reported {\te}({\oiii}) =
\mbox{11\,553~$\pm$~1579\,K},
{\te}({\oii}) = \mbox{11\,983~$\pm$~770\,K}\footnote{However,
the auroral {\oii} lines are out of their spectrum taken in 2011.},
{\te}({\nii}) = \mbox{11\,066~$\pm$~1752\,K}, 
{\te}({\siii}) =
\mbox{11\,831~$\pm$~2286\,K}\footnote{The nebular {\siii} lines are
out of their spectrum, too.}, and {\Ne}({\sii}) =
 \mbox{8740~$\pm$~7701\,cm$^{-3}$}, respectively. The difference
between their {\te}({\oiii}) and ours is due to the
{\oiii}\,4363\,{\AA} intensity (see appendix Table~\ref{T:comp_obs}). Under a
constant  {\Ne}, the {\te}({\oiii}) becomes higher as the {\oiii}
\mbox{$I$(4959/5007\,{\AA})/$I$(4363\,{\AA})} ratio becomes lower.
The {\nii} {\Ne}-{\te} curve in Fig.~\ref{F:diagno} suggests that
the discrepancy in {\te}({\nii}) could be due to the difference in adopted {\Ne}.

\subsection{Ionic abundance derivations \label{S:ionic}}

 \begin{table*}[t!]
\caption{The ionic abundances derived using CELs. \label{T:CEL}}
\tablecolumns{10}
\renewcommand{\arraystretch}{0.95}
  \footnotesize
\centering
\tablewidth{\textwidth}
\begin{tabular}{@{}c@{\hspace{6pt}}c@{\hspace{6pt}}c@{\hspace{6pt}}D{p}{\pm}{-1}@{\hspace{6pt}}D{p}{\pm}{-1}@{\hspace{6pt}}c@{\hspace{6pt}}c@{\hspace{6pt}}c@{\hspace{6pt}}D{p}{\pm}{-1}@{\hspace{6pt}}D{p}{\pm}{-1}@{}}
\hline\hline
Elem. &Ion        &$\lambda_{\rm lab.}$ &\multicolumn{1}{c}{$I$($\lambda$)}   &\multicolumn{1}{c}{$n$(X$^{\rm m+}$)/$n$(H$^{+}$)}&
Elem. &Ion        &$\lambda_{\rm lab.}$ &\multicolumn{1}{c}{$I$($\lambda$)}   &\multicolumn{1}{c}{$n$(X$^{\rm m+}$)/$n$(H$^{+}$)}\\
(X)   &(X$^{\rm m+}$) &             &\multicolumn{1}{c}{($I$({\hb}) = 100)} &
&(X)   &(X$^{\rm m+}$) &              &\multicolumn{1}{c}{($I$({\hb}) = 100)}\\
(1)&(2)&(3)&\multicolumn{1}{c}{(4)}&\multicolumn{1}{c}{(5)}&(6)&(7)&(8)&\multicolumn{1}{c}{(9)}&\multicolumn{1}{c}{(10)}\\
\hline
N(CEL) & N$^{0}$ & 5197.90\,{\AA} & 3.63(-1) ~p~ 4.38(-3) & 1.11(-6) ~p~ 3.46(-8) & S & S$^{+}$ & 4068.60\,{\AA} & 4.47(0) ~p~ 8.61(-2) & 1.13(-6) ~p~ 1.37(-7) \\
 &  & 5200.26\,{\AA} & 2.28(-1) ~p~ 3.59(-3) & 1.07(-6) ~p~ 1.82(-8) &  &  & 4076.35\,{\AA} & 1.51(0) ~p~ 2.90(-2) & 1.17(-6) ~p~ 1.43(-7) \\
 &  &  &\multicolumn{1}{c}{\bf Average}     & {\bf 1.09(-6)} ~p~ {\bf 2.83(-8)} &  &  & 6716.44\,{\AA} & 6.07(0) ~p~ 1.54(-1) & 9.55(-7) ~p~ 1.57(-7) \\
 & N$^{+}$ & 5754.64\,{\AA} & 2.55(0) ~p~ 3.87(-2) & 3.61(-5) ~p~ 2.41(-6) &  &  & 6730.81\,{\AA} & 1.25(+1) ~p~ 3.19(-1) & 1.07(-6) ~p~ 1.13(-7) \\
 &  & 6548.04\,{\AA} & 4.10(+1) ~p~ 9.67(-1) & 3.60(-5) ~p~ 1.24(-6) &
       &  &  &\multicolumn{1}{c}{\bf Average}     & {\bf 1.06(-6)} ~p~ {\bf 1.30(-7)} \\
 &  & 6583.46\,{\AA} & 1.21(+2) ~p~ 2.91(0) & 3.60(-5) ~p~ 1.25(-6) &  & S$^{2+}$ & 6312.10\,{\AA} & ~1.03(0) ~p~ 2.19(-2) & 6.19(-6) ~p~ 9.21(-7) \\
 &  &  &\multicolumn{1}{c}{\bf Average}     & {\bf 3.60(-5)} ~p~ {\bf 1.27(-6)} &  &  & 9068.60\,{\AA} & 1.26(+1) ~p~ 5.96(-1) & 4.90(-6) ~p~ 3.88(-7) \\
 &  &  &\multicolumn{1}{c}{ICF(N(CEL))}    & 3.09 ~p~ 0.17 &  &  & 18.71\,{\micron} & 1.39(+1) ~p~ 4.65(-1) & 5.60(-6) ~p~ 8.07(-7) \\
 &  &  &     & {\bf 1.11(-4)} ~p~ {\bf 7.39(-6)} &  &  & 33.48\,{\micron} & 3.52(0) ~p~ 2.72(-1) & 5.63(-6) ~p~ 1.31(-6) \\
O(CEL) & O$^{0}$ & 5577.34\,{\AA} & 2.19(-1) ~p~ 4.49(-3) & 6.06(-5) ~p~
   3.96(-6) &  &  &  & \multicolumn{1}{c}{\bf Average}    & {\bf
				     5.34(-6)} ~p~ {\bf 6.98(-7)} \\
 &  & 6300.30\,{\AA} & 1.61(+1) ~p~ 3.39(-1) & 6.06(-5) ~p~ 2.48(-6) &  & S$^{3+}$ & 10.51\,{\micron} & 7.76(0) ~p~ 2.66(-1) & 4.18(-7) ~p~ 6.52(-8) \\
 &  & 6363.78\,{\AA} & 5.10(0) ~p~ 1.11(-1) & 6.00(-5) ~p~ 2.46(-6) &  &  &  &\multicolumn{1}{c}{ ICF(S)}    &\multicolumn{1}{c}{\,\,1.00}     \\
 &  &  & \multicolumn{1}{c}{\bf Average}    & {\bf 6.05(-5)} ~p~ {\bf 2.49(-6)} &  &  &  &
     & {\bf 6.82(-6)} ~p~ {\bf 7.13(-7)} \\
 & O$^{+}$ & 3726.03\,{\AA} & 1.01(+2) ~p~ 2.62(0) & 2.67(-4) ~p~ 1.08(-5) & Ar & Ar$^{+}$ & 6.99\,{\micron} & 7.43(0) ~p~ 2.53(-1) & 7.15(-7) ~p~ 2.48(-8) \\
 &  & 3728.81\,{\AA} & 3.76(+1) ~p~ 9.79(-1) & 2.56(-4) ~p~ 9.88(-6) &  & Ar$^{2+}$ & 5191.82\,{\AA} & 6.41(-2) ~p~ 3.11(-3) & 1.94(-6) ~p~ 3.27(-7) \\
 &  & 7320/7330\,{\AA} & 3.18(+1) ~p~ 5.56(-1) & 2.98(-4) ~p~ 2.32(-5) &  &  & 7135.80\,{\AA} & 1.08(+1) ~p~ 3.22(-1) & 1.47(-6) ~p~ 1.16(-7) \\
 &  &  &\multicolumn{1}{c}{\bf Average}     & {\bf 2.70(-4)} ~p~ {\bf 1.29(-5)} &  &  & 7751.10\,{\AA} & 2.63(0) ~p~ 9.57(-2) & 1.50(-6) ~p~ 1.23(-7) \\
 & O$^{2+}$ & 4363.21\,{\AA} & 2.46(0) ~p~ 2.96(-2) & 1.88(-4) ~p~ 9.02(-6) &  &  & 9.01\,{\micron} & 1.30(+1) ~p~ 4.71(-1) & 1.70(-6) ~p~ 6.56(-8) \\
 &  & 4931.23\,{\AA} & ~5.46(-2) ~p~ 2.92(-3) & 1.80(-4) ~p~ 9.67(-6) &
       &  &  &\multicolumn{1}{c}{\bf Average}     & {\bf 1.59(-6)} ~p~ {\bf 9.25(-8)} \\
 &  & 4958.91\,{\AA} & 1.46(+2) ~p~ 3.80(-1) & 1.87(-4) ~p~ 1.26(-6) &  & Ar$^{3+}$ & 4711.37\,{\AA} & 3.39(-2) ~p~ 3.05(-3) & 2.12(-8) ~p~ 2.29(-9) \\
 &  &  &\multicolumn{1}{c}{\bf Average}     & {\bf 1.87(-4)} ~p~ {\bf 1.39(-6)} &  &  & 4740.16\,{\AA} & 6.95(-2) ~p~ 1.72(-3) & 2.09(-8) ~p~ 8.27(-10) \\
 &  &  &\multicolumn{1}{c}{ ICF(O(CEL))}    &\multicolumn{1}{c}{\,\,1.00}   &  &  &  &\multicolumn{1}{c}{\bf
				 Average}     & {\bf 2.10(-8)} ~p~ {\bf 1.31(-9)} \\
 &  &  &     & {\bf 4.57(-4)} ~p~ {\bf 1.30(-5)} &  &  &  &\multicolumn{1}{c}{ ICF(Ar)}     &\multicolumn{1}{c}{\,\,1.00 }   \\
Ne & Ne$^{+}$ & 12.81\,{\micron} & 4.67(+1) ~p~ 1.55(0) & 7.23(-5) ~p~
   2.45(-6) &  &  &  &     & {\bf 2.32(-6)} ~p~ {\bf 9.57(-8)} \\
 & Ne$^{2+}$ & 3869.06\,{\AA} & 4.03(+1) ~p~ 9.46(-1) & 8.77(-5) ~p~ 4.07(-6) & Fe & Fe$^{2+}$ & 4658.05\,{\AA} & 1.11(-1) ~p~ 2.50(-3) & 6.71(-8) ~p~ 2.80(-9) \\
 &  & 3967.79\,{\AA} & 1.00(+1) ~p~ 2.16(-1) & 7.22(-5) ~p~ 3.29(-6) &  &  & 4701.53\,{\AA} & 4.65(-2) ~p~ 3.17(-3) & 7.17(-8) ~p~ 5.65(-9) \\
 &  & 15.56\,{\micron} & 9.67(+1) ~p~ 3.19(0) & 8.38(-5) ~p~ 4.54(-6) &  &  & 4733.91\,{\AA} & 2.58(-2) ~p~ 2.74(-3) & 8.82(-8) ~p~ 1.01(-8) \\
 &  & 36.02\,{\micron} & 6.18(0) ~p~ 3.23(-1) & 8.57(-5) ~p~ 1.01(-5) &  &  & 4754.69\,{\AA} & 2.86(-2) ~p~ 3.64(-3) & 9.18(-8) ~p~ 1.21(-8) \\
 &  &  & \multicolumn{1}{c}{\bf Average}    & {\bf 8.41(-5)} ~p~ {\bf 4.56(-6)} &  &  & 5270.40\,{\AA} & 5.89(-2) ~p~ 2.40(-3) & 6.76(-8) ~p~ 3.48(-9) \\
 &  &  & \multicolumn{1}{c}{ICF(Ne)}    & \multicolumn{1}{c}{\,\,1.00}   &  &  &   &\multicolumn{1}{c}{\bf Average} & {\bf 7.26(-8)} ~p~ {\bf 5.11(-9)} \\
 &  &  &     & {\bf 1.56(-4)} ~p~ {\bf 4.77(-6)} &  &  &  &\multicolumn{1}{c}{ ICF(Fe)}   & 2.30 ~p~ 0.14 \\
Cl & Cl$^{+}$ & 8578.69\,{\AA} & 3.04(-1) ~p~ 1.35(-2) & 1.64(-8) ~p~
   8.17(-10) &  &  &  &    & {\bf 1.67(-7)} ~p~ {\bf 1.57(-8)} \\
 &  & 9123.60\,{\AA} & 1.03(-1) ~p~ 5.46(-3) & 2.12(-8) ~p~ 1.23(-9) &  &  &  &    &    \\
 &  &  &\multicolumn{1}{c}{\bf Average}     & {\bf 1.76(-8)} ~p~ {\bf 9.22(-10)} &  &  &  &    &    \\
 & Cl$^{2+}$ & 5517.72\,{\AA} & 1.17(-1) ~p~ 3.39(-3) & 1.02(-7) ~p~ 3.71(-8) &  &  &  &    &    \\
 &  & 5537.89\,{\AA} & 2.40(-1) ~p~ 4.59(-3) & 1.02(-7) ~p~ 3.93(-8) &  &  &  &    &    \\
 &  & 8434.00\,{\AA} & 7.76(-3) ~p~ 1.23(-3) & 1.18(-7) ~p~ 7.51(-8) &  &  &  &    &    \\
 &  & 8500.20\,{\AA} & 9.76(-3) ~p~ 3.70(-3) & 1.18(-7) ~p~ 8.55(-8) &  &  &  &    &    \\
 &  &  &\multicolumn{1}{c}{\bf Average}     & {\bf 1.03(-7)} ~p~ {\bf 4.06(-8)} &  &  &  &    &    \\
 &  &  &\multicolumn{1}{c}{ ICF(Cl) }   & 1.01 ~p~ 0.06 &  &  &  &    &    \\
 &  &  & & {\bf 1.21(-7)} ~p~ {\bf 4.16(-8)} &  &  &  &    &    \\
\hline
\end{tabular}
 \end{table*}

 \begin{table}[t!]
  \caption{The ionic abundances derived using RLs. \label{T:RL}}
  \renewcommand{\arraystretch}{0.95}
\footnotesize
\centering
\tablecolumns{5}
\tablewidth{\columnwidth}
\begin{tabular}{@{}ccc@{\hspace{4pt}}D{p}{\pm}{-1}@{\hspace{4pt}}D{p}{\pm}{-1}@{}}
\hline\hline
Elem. &Ion        &$\lambda_{\rm lab.}$ &\multicolumn{1}{c}{$I$($\lambda$)}   &\multicolumn{1}{c}{$n$(X$^{\rm m+}$)/$n$(H$^{+}$)}\\
(X)   &(X$^{\rm m+}$) &({\AA})              &\multicolumn{1}{c}{($I$({\hb}) = 100)} &\\ 
\hline
He & He$^{+}$ & 4120.81 & 1.77(-1) ~p~ 4.80(-3) & 9.95(-2) ~p~ 5.32(-3) \\ 
 &  & 4387.93 & 4.39(-1) ~p~ 6.42(-3) & 7.18(-2) ~p~ 4.73(-3) \\ 
 &  & 4437.55 & 6.64(-2) ~p~ 4.31(-3) & 8.53(-2) ~p~ 6.61(-3) \\ 
 &  & 4471.47 & 4.63(0) ~p~ 4.85(-2) & 9.34(-2) ~p~ 5.22(-3) \\ 
 &  & 4713.22 & 6.68(-1) ~p~ 6.70(-3) & 1.14(-1) ~p~ 9.20(-3) \\ 
 &  & 4921.93 & 1.21(0) ~p~ 4.45(-3) & 9.05(-2) ~p~ 4.65(-3) \\ 
 &  & 5015.68 & 2.16(0) ~p~ 1.19(-2) & 7.69(-2) ~p~ 3.63(-3) \\ 
 &  & 5047.74 & 1.68(-1) ~p~ 3.09(-3) & 8.24(-2) ~p~ 4.52(-3) \\ 
 &  & 5875.60 & 1.47(+1) ~p~ 2.62(-1) & 1.02(-1) ~p~ 6.54(-3) \\ 
 &  & 6678.15 & 3.97(0) ~p~ 9.94(-2) & 9.79(-2) ~p~ 5.74(-3) \\ 
 &  & 7281.35 & 7.26(-1) ~p~ 2.30(-2) & 8.37(-2) ~p~ 4.43(-3) \\ 
 &  &  &\multicolumn{1}{c}{\bf Average} &    {\bf 9.69(-2)} ~p~ {\bf 5.88(-3)} \\ 
 &  &  &\multicolumn{1}{c}{ICF(He)} &  1.00    \\ 
 &  &  & &    {\bf 9.69(-2)} ~p~ {\bf 5.88(-3)} \\ 
C(RL) & C$^{2+}$ & 4267.18 & 1.03(-1) ~p~ 4.54(-3) & 9.66(-5) ~p~ 3.06(-5) \\ 
      &          & 6578.05 & 5.05(-2) ~p~ 3.71(-3) & 9.84(-5) ~p~ 3.88(-5) \\ 
      &          &         &\multicolumn{1}{c}{\bf Average}           &
		 {\bf 9.72(-5)} ~p~ {\bf 3.33(-5)} \\ 
      &          &         &\multicolumn{1}{c}{ICF(C(RL))}              & 1.48 ~p~ 0.22 \\ 
      &          &         &                         & {\bf 1.44(-4)}
		 ~p~ {\bf 5.38(-5)} \\ 
N(RL) & N$^{2+}$ & 4630.54 & 2.02(-2) ~p~ 2.59(-3) & 9.27(-5) ~p~ 4.21(-5) \\ 
 &  & 5679.56 & 1.69(-2) ~p~ 2.30(-3) & 4.23(-5) ~p~ 1.93(-5) \\ 
    &         & &\multicolumn{1}{c}{\bf Average}           &  {\bf 6.97(-5)} ~p~ {\bf 3.17(-5)} \\ 
    &         & &\multicolumn{1}{c}{ICF(N(RL))}              & 1.48 ~p~ 0.22 \\ 
    &  &      &                         & {\bf 1.03(-4)} ~p~ {\bf 4.94(-5)} \\ 
O(RL) & O$^{2+}$ & 4069.62 & 2.65(-2) ~p~ 3.24(-3) & 2.78(-4) ~p~ 7.04(-5) \\ 
 &  & 4069.88 & 3.84(-2) ~p~ 4.55(-3) & 2.52(-4) ~p~ 6.61(-5) \\ 
 &  & 4072.15 & 5.68(-2) ~p~ 2.51(-3) & 2.34(-4) ~p~ 5.45(-5) \\ 
 &  & 4075.86 & 7.55(-2) ~p~ 5.04(-3) & 2.26(-4) ~p~ 5.34(-5) \\ 
 &  & 4104.99 & 4.05(-2) ~p~ 4.28(-3) & 3.64(-4) ~p~ 9.07(-5) \\ 
 &  & 4153.30 & 3.10(-2) ~p~ 1.60(-3) & 4.04(-4) ~p~ 9.69(-5) \\ 
 &  & 4349.43 & 3.34(-2) ~p~ 2.39(-3) & 1.70(-4) ~p~ 3.92(-5) \\ 
 &  & 4366.90 & 3.41(-2) ~p~ 2.60(-3) & 4.48(-4) ~p~ 1.07(-4) \\ 
 &  & 4638.86 & 3.44(-2) ~p~ 3.53(-3) & 3.06(-4) ~p~ 7.73(-5) \\ 
 &  & 4641.81 & 6.37(-2) ~p~ 2.21(-3) & 2.38(-4) ~p~ 5.35(-5) \\ 
 &  & 4649.13 & 1.03(-1) ~p~ 2.72(-3) & 2.13(-4) ~p~ 4.79(-5) \\ 
 &  & 4650.84 & 3.42(-2) ~p~ 1.93(-3) & 3.28(-4) ~p~ 7.57(-5) \\ 
 &  & 4661.63 & 4.79(-2) ~p~ 2.17(-3) & 3.83(-4) ~p~ 8.68(-5) \\ 
 &  & 4676.23 & 3.19(-2) ~p~ 5.12(-3) & 3.34(-4) ~p~ 9.09(-5) \\ 
 &  &         &\multicolumn{1}{c}{\bf Average}          & {\bf 2.82(-4)} ~p~ {\bf 6.73(-5)} \\ 
 &  &         &\multicolumn{1}{c}{ICF(O(RL))}             & 2.45 ~p~ 0.07 \\ 
 &  &         &                        & {\bf 6.89(-4)} ~p~ {\bf 1.66(-4)} \\ 
\hline
\end{tabular}
 \end{table}

\begin{table}[t!]
 \caption{Comparison of the ionic abundances in 2006 by us and 2011 by 
 \citet{Arkhipova:2013aa}.  \label{T:comp_abund}}
\renewcommand{\arraystretch}{0.95}
 \centering
 \footnotesize
\begin{tabular}{@{}lcc@{}}
\hline\hline
Ion (X$^{\rm m+}$)       &$n$(X$^{\rm m+}$)/$n$(H$^{+}$) in
     2006&$n$(X$^{\rm m+}$)/$n$(H$^{+}$) in 2011\\
\hline
He$^{+}$      &9.69(--2) $\pm$ 5.88(--3)  &  9.70(--2) $\pm$ 8.00(--3)\\
N$^{+}$       &3.60(--5) $\pm$ 1.27(--6)  & 5.81(--5) $\pm$ 2.31(--5)\\
O$^{+}$       &2.70(--4) $\pm$ 1.29(--5)  &9.10(--5) $\pm$ 7.16(--5)\\
O$^{2+}$(CEL) &1.87(--4) $\pm$ 1.39(--6)  &1.01(--4) $\pm$ 4.26(--5)\\
Ne$^{2+}$     &8.41(--5) $\pm$ 4.56(--6)  &3.46(--5) $\pm$ 1.76(--5)\\
S$^{+}$       &1.06(--6) $\pm$ 1.30(--7)  &9.34(--7) $\pm$ 5.70(--7)\\
S$^{2+}$      &5.34(--6) $\pm$ 6.98(--7)  &1.43(--6) $\pm$ 1.13(--6)\\
Ar$^{2+}$     &1.59(--6) $\pm$ 9.25(--8)  &1.00(--6) $\pm$ 4.26(--7)\\
\hline
\end{tabular}
 \end{table}

In appendix Table~\ref{T:nete}, we list {\Ne} and {\te} adopted for 
calculating each ionic abundance. We determined these values
by referring to the {\Ne}-{\te} diagram and taking the ionization
potential (IP)
of the targeting ion into account. We calculated the CEL
ionic abundances by solving an equation of population
at multiple energy levels (from two energy
levels for Ne$^{+}$ and Ar$^{+}$ and 33 levels for Fe$^{2+}$)
under the listed {\Ne} and {\te}.
We adopted a constant {\Ne} = 10$^{4}$\,cm$^{-3}$ and
the average {\te}({\hei}) 8160\,K to calculate the He$^{+}$.
For the RL C$^{2+}$, N$^{2+}$, and O$^{2+}$,
we adopted {\Ne} = 10$^{4}$\,cm$^{-3}$ and {\te}(PJ).

We summarize the resultant CEL and RL ionic abundances in 
Tables~\ref{T:CEL} and \ref{T:RL}, respectively.
When we detected two or more lines of a target ion, we derived each
ionic abundance using each line intensity. Then, we adopted the
weight-average value as the representative ionic abundance
as listed in the last line of each ion by boldface. We give 
the 1-$\sigma$ uncertainty of each ionic abundance, which accounts 
for the uncertainties of line fluxes (including $c$({\hb})
uncertainty), {\te}, and {\Ne}.

The CEL abundances calculated using the optical lines are well consistent
with ones using mid-IR fine-structure lines, indicating that the
calculated CEL
ionic abundances are the results based on proper selections of
{\te} in particular and accurate scaling flux of the \emph{Spitzer}/IRS spectrum.

We obtained the RL N$^{2+}$ and O$^{2+}$ in this PN for the first time.
The higher
multiplet lines are in general insensitive to Case A or Case B assumptions and
reliable because these lines are less affected by both resonance fluorescence
by starlight and recombination from higher terms. The consistency between
the RL C$^{2+}$ abundance by the multiple V6 4267.18\,{\AA} line and by
the V2 6578.05\,{\AA} line indicates that the RL C$^{2+}$ from both
lines can be reliable. We can have the similar conclusion for the RL
O$^{2+}$ and N$^{2+}$ abundances. The RL O$^{2+}$ abundances are well consistent among
the O\,{\sc ii} V1 4638/42/49/51/62/76\,{\AA}, V2 4349/67\,{\AA}, V10
4069.6/69.9/72/76\,{\AA}, V19 4153\,{\AA}, and V20 4105\,{\AA}
lines. The RL N$^{2+}$ abundances are derived using the V3 5679\,{\AA}
and 4631\,{\AA} lines.

As compared in Table~\ref{T:comp_abund}, our ionic abundances agree with
\citet{Arkhipova:2013aa}. However, we found the obvious
discrepancies in the Ne$^{2+}$ and S$^{2+}$.
Their S$^{2+}$ seems to be derived using the auroral
line {\siii}\,6312\,{\AA}. Although they did not report the detection of any {\neiii} lines
in their spectrum taken in 2011, we assume that they derived the
Ne$^{2+}$ using nebular {\neiii} lines.
The Ne$^{2+}$ and S$^{2+}$ differences between
\citet{Arkhipova:2013aa} and us are due to the adopted {\te}.
We stress that our adopted {\te} for the Ne$^{2+}$ and S$^{2+}$ is
determined using the {\neiii} and {\siii} fine-structure, nebular,
and auroral lines. For instance, if we adopt
their {\te}({\oiii}) = 11\,553\,K to calculate the Ne$^{2+}$ using the nebular {\neiii} lines, the volume
emissivities of these {\neiii} lines become 2.66 times higher than those in our
adopted {\te} = 8560\,K. Accordingly, the Ne$^{2+}$ is down to 3.18(--5), which
is consistent with \citet{Arkhipova:2013aa}. However, since the emissivities of
the fine-structure {\neiii} lines do not largely change even in both
8560\,K by ours and 11\,553\,K by \citet{Arkhipova:2013aa}, the
Ne$^{2+}$ abundances using
the fine-structure {\neiii} lines keep 8.38(--5)
(from the {\neiii}\,15.56\,{\micron} line) and 8.57(--5) (from the
{\neiii}\,36.02\,{\micron} line). That is, we find out the {\it
spurious} Ne$^{2+}$ derivation discrepancy between the nebular and the fine-structure lines. We
confirmed that the similar conclusion can apply for the
S$^{2+}$. Thus, if the nebula condition is in a steady state and had not dramatically
changed in 2006-2011, we can conclude that our Ne$^{2+}$ and
S$^{2+}$ are more reliable.

\subsection{Elemental abundance derivations using the ICFs \label{S:elementICF}}

 \begin{table}[t!]
  \caption{Elemental abundances. The C(CEL) is an {\it expected} value
 estimated by adopting the RL C/O ratio. \label{T:abund}}
 \centering
\footnotesize
\renewcommand{\arraystretch}{0.95}
 \tablewidth{\textwidth}
\begin{tabular}{@{}lD{p}{\pm}{-1}D{p}{\pm}{-1}D{p}{\pm}{-1}@{}}
\hline\hline
Elem. (X)   &\multicolumn{1}{c}{$n$(X)/$n$(H)}                 &\multicolumn1c{$\epsilon$(X)}
 &\multicolumn{1}{c}{[X/H]}\\
\hline
He &9.69(-2)~p~5.88(-3) &10.99~p~0.03 &+0.06~p~0.03\\
C(RL)  &1.44(-4)~p~5.38(-5) &8.16~p~0.16 &-0.23~p~0.17\\
C(CEL) &9.54(-5)~p~4.26(-5) &7.98~p~0.19 &-0.40~p~0.20\\
N(RL) &1.03(-4)~p~7.39(-6) &8.01~p~0.03 &+0.15~p~0.12\\
N(CEL) &1.11(-4)~p~7.39(-6) &8.05~p~0.03 &+0.19~p~0.12\\
O(RL) &6.89(-4)~p~1.66(-4) &8.84~p~0.10 &+0.11~p~0.13\\
O(CEL) &4.57(-4)~p~1.30(-5) &8.66~p~0.01 &-0.07~p~0.07\\
Ne &1.56(-4)~p~4.77(-6) &8.19~p~0.01 &+0.14~p~0.10\\
S &6.82(-6)~p~7.13(-7) &6.83~p~0.05 &-0.33~p~0.05\\
Cl &1.21(-7)~p~4.16(-8) &5.08~p~0.15 &-0.17~p~0.16\\
Ar &2.32(-6)~p~9.57(-8) &6.37~p~0.02 &-0.13~p~0.10\\
Fe &1.67(-7)~p~1.57(-8) &5.22~p~0.04 &-2.24~p~0.09\\
\hline
\end{tabular}
 \end{table}

By introducing the ionization correction factor (ICF), we inferred 
the nebular abundances from their ionic abundances. We calculated
these ICF(X)s derived based on the fraction of the observed ionic
abundances with similar ionization potentials to the target element.
The ICF(X) of element X is listed in the
last line of each element of Tables~\ref{T:CEL} and
\ref{T:RL}. The abundance of the element X $n$(X)/$n$(H) corresponds
to the value derived from the \mbox{ICF(X)~$\cdot$~$\Sigma_{\rm m=1}$$n$(X$^{\rm m+}$)/$n$(H$^{+}$)}. 
We will compare these ICFs(X) based on IPs with those calculated by Cloudy photoionization model later.

As shown in \S \ref{S:ionic}, we obtained the
O(CEL), Ne, S, and Ar ionic abundances in various ionization stages.
Thus, for these elements, we can adopt the ICF(X) = 1.0. We adopted the
ICF(He) = 1.0 because we did not detect the nebular He\,{\sc ii} lines.
Assuming that N corresponds to the sum of the N$^{+}$
and N$^{2+}$, we recovered the unobserved N$^{2+}$(CEL) using the ICF(N)
proposed by \cite{Delgado-Inglada:2014ab}. Then, using the ICF(N) for
the N(CEL), we determined the ICF(N(RL)). Since the IPs in both C and N
ions are similar, we assumed that ICF(C(RL)) is as same as the
ICF(N(RL)). The ICF(O(RL)) corresponds to the O(CEL)/O$^{2+}$(CEL). The
ICF(Cl) corresponds to the \mbox{Ar/(Ar$^{+}$~+~Ar$^{2+}$)} ratio.
For the ICF(Fe), we adopted equation (3) of
\cite{Delgado-Inglada:2014aa}.

In Table~\ref{T:abund}, we summarize the resultant elemental abundances
derived by introducing the ICFs. The value $\epsilon$(X) in the third column
is \mbox{12~+~$\log_{10}$$n$(X)/$n$(H)}.
The value in the last column is the relative abundance to the
Sun. We referred the solar abundance by \citet{Asplund:2009aa}.
Our work improved nebular elemental abundances calculated by the
pioneering work of \citet{Parthasarathy:1993aa} and a recent comprehensive
study of \citet{Arkhipova:2013aa}.

Using the RL C, N, and O, we derived the C/O
and the N/O ratios using the same type of emission lines, i.e., RLs.
These ratios are important proofs of the initial mass of the central
star. In Table~\ref{T:abund}, we
list an {\it expected } C(CEL) based on the assumption that the RL C/O
ratio (\mbox{0.21~$\pm$~0.09}) is consistent with the CEL C/O ratio.

The RL C/O ratio indicates that Hen3-1357 is an O-rich PN, which is also
supported by the detection of the amorphous silicate features.
The average of the logarithmic difference between the nebular and
solar abundances of S, Cl, and Ar $\langle$[S,Cl,Ar/H]$\rangle$ =
\mbox{--0.21~$\pm$~0.10}
 indicates that this PN is about a half of solar metallicity (0.6\,$Z_{\sun}$). In the
Milky Way chemical evolution in such metallicity, the [Fe/H] should be
comparable to the [$\alpha$/H]. The expected [Fe/H] is --0.23 from  the average
[S,Ar/H] $\sim$--0.23 if all the Fe-atoms are in gas phase and are
not captured by any dust grains. However, the
observed [Fe/H] is much smaller than the expected [Fe/H] value. Thus, the largely depleted [Fe/H]
suggests that over 99\,$\%$ of the Fe-atoms in the nebula would be
locked within silicate grains.

\subsection{Abundance discrepancy of the C$^{2+}$, N$^{2+}$, and O$^{2+}$}

One of the long-standing problems in PN abundances is
that the RL C, N, O, and Ne ionic abundances are in general larger than
those CEL ones. Several explanations for the abundance discrepancy have been proposed,
e.g., temperature fluctuation, high density clumps, and
cold hydrogen-deficient components \citep[see, e.g., a  review
by][]{Liu:2006aa}. There might be a possible link between the binary
central star and the abundance discrepancy, as recently proposed
by \citet{Jones:2016aa}. In Hen3-1357, the abundance discrepancy factor
(ADF) defined as the ratio of the RL to the CEL ionic abundance is
\mbox{1.51~$\pm$~0.36} in the O$^{2+}$, which is lower than a typical value $\sim$2.0
\citep{Liu:2006aa}. Such degree of the O$^{2+}$ discrepancy
can be explained by introducing temperature fluctuation model proposed
by \citet{Peimbert:1967aa}.

\cite{Parthasarathy:1993aa} and \cite{Feibelman:1995aa} showed the \emph{IUE} UV-spectrum taken on 1992
April 23 (IUE Program ID: NA108, PI: S.R.~Pottasch,
Data-ID: 44459). There, we can see the CEL C\,{\sc iii}$]$\,1906/09\,{\AA} 
and N\,{\sc iii}$]$\,1744-54\,{\AA} lines. Although
\citet{1995A&A...300L..25P} gave these line fluxes, the CEL C$^{2+}$ and
N$^{2+}$ have never been calculated so far.
It would be of interest to estimate the ADF(C$^{2+}$) 
because the RL C$^{2+}$ of \mbox{6.91(--5)~$\pm$~1.48(--6)} 
using the C\,{\sc ii}\,4267\,{\AA} line detected in the spectrum taken in
1992 was calculated by \citet{Arkhipova:2013aa}. We download the
processed SWP44459 dataset from Multimission Archive at STScI (MAST), we measured the fluxes
of the C\,{\sc iii}$]$\,1906/09\,{\AA} and N\,{\sc iii}$]$\,1744-54\,{\AA} lines, and calculated the CEL C$^{2+}$
= \mbox{9.53(--5)~$\pm$~1.19(--6)} and the CEL N$^{2+}$ =
\mbox{5.49(--5)~$\pm$~5.32(--6)} using the $F$({\hb}), 
$E$($B$-$V$), {\te}, and {\Ne} reported by \cite{Parthasarathy:1993aa}. 
The ADF(C$^{2+}$) of \mbox{1.24~$\pm$~0.27} in 1992 is similar
to our ADF(O) measured in 2006. This might be applied even for ADF(N);
the ratio of the RL N$^{2+}$ in 2006 to the CEL N$^{2}$ in
1992 is \mbox{1.27~$\pm$~0.59}.

Based on our analysis, we conclude that the ADF(C/N/O) would be $<$~2.

\subsection{Comparison with AGB nucleosynthesis models \label{S:AGB}}

The He/C/N/O/Ne/S abundances are close to the
AGB star nucleosynthesis model predictions by \cite{Karakas:2010aa} for
an initially
1.0, 1.25, and 1.5\,M$_{\sun}$ stars with \mbox{$Z$ = 0.008}
(0.4\,$Z_{\sun}$). The 1.0\,$M_{\sun}$ model predicts
$\epsilon$(He):10.99, $\epsilon$(C):8.09, $\epsilon$(N):7.81,
$\epsilon$(O):8.53, $\epsilon$(Ne):7.69, and $\epsilon$(S):7.00. The
differences among these
models are $\epsilon$(C) and $\epsilon$(N); $\epsilon$(C):8.04 and
$\epsilon$(N):7.90 in the 1.25\,$M_{\sun}$ model, and
$\epsilon$(C):8.12 and $\epsilon$(N):7.95 in the 1.5\,$M_{\sun}$
model. The predicted final core-mass is 0.58\,$M_{\sun}$ in an
initially 1.0\,$M_{\sun}$ star to 0.63\,$M_{\sun}$ in
a 1.5\,$M_{\sun}$ star.

According to current stellar models for low-mass AGB stars,
partial mixing of the bottom of the H-rich convective envelope
into the outermost region of the $^{12}$C-rich intershell layer
leads to the synthesis of extra $^{13}$C and $^{14}$N at the end
of each third dredge-up (TDU). During He-burning, $^{14}$N captures two $\alpha$
particles, and $^{22}$Ne are produced. $^{20}$Ne is the most
abundant, and it is not altered significantly by H- or He-burning.
The $\epsilon$(Ne) discrepancy between the observation (8.19)
and the model prediction (7.69) might be due to an increase of
$^{22}$Ne. The models for the 1.0, 1.25, and 1.5\,$M_{\sun}$
stars with $Z$ = 0.008 by \cite{Karakas:2010aa} do not predict TDUs
and do not include such partial mixing zone (PMZ). Note that
PMZ is not well justified yet. The Ne abundance in Hen3-1357 suggests
that the progenitor might have formed PMZ and extra $^{22}$Ne
and Ne might be conveyed to the stellar surface by unexpected mechanisms,
e.g., very few TDUs or LTPs. Otherwise, we might interpret that the
$\epsilon$(Ne) discrepancy between the observation
and the model prediction is due to the errors in the atomic
data of Ne$^{+,2+}$.

From chemical abundance analysis, we can
conclude that the progenitor mass could be 1.0-1.5\,$M_{\sun}$ if Hen3-1357
has evolved from a star with the initial $Z$$\sim$0.008.

\section{Photoionization model with Cloudy}

We construct the self-consistent photoionization model
using Cloudy to reproduce
all the observed quantities.

The characteristics of the CSPN are critical in the photoionization models
because the X-ray to UV wavelength radiation from CSPN determines the ionization
structure of the nebula and surrounding ISM and is the ionizing and
heating source of gas and dust grains. The distance is necessary for the comparison between the
model and the observed fluxes/flux densities/nebula size. In
\S \ref{S:cspn} and \ref{S:cspn2}, therefore, we try to determine parameters of
the CSPN and the distance.

The empirically derived quantities of the nebula and the mid-IR SED
provide the input parameters of the nebula and dust grain:
$\epsilon$(X), geometry, the {\hb} flux of
the entire nebula, hydrogen density radial profile ($n_{\rm H}$) of the nebula,
filling factor ($f$), and type of dust grain.
The band flux densities/fluxes, gas emission-line fluxes, and the SED
from the UV to far-IR provide constraints in the iterative
fitting of the model parameters. In \S \ref{S:modelneb}, we
explain the input parameters. Finally, we show the modeling result in \S \ref{S:resmodel}.

\subsection{Flux density of the CSPN's SED \label{S:cspn}}

\begin{figure}[t!]
   \includegraphics[width=\columnwidth,clip]{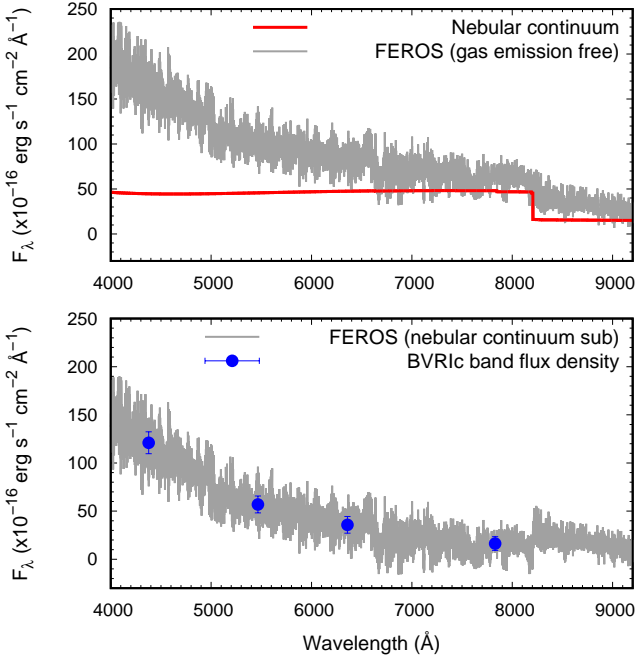}
   \caption{({\it upper panel}) The gas-emission line free FEROS
   spectrum of Hen3-1357 (de-reddened, grey line) and the synthesized
   nebular continuum by {\sc Nebcont} (red line) in the range from 4000
 to 9200\,{\AA}. ({\it lower panel})
   The nebular continuum subtracted FEROS spectrum (grey line) and the
 de-reddened $BVRIc$ band flux densities (blue circles)
 based on this residual FEROS spectrum. \label{S:feros2}}
\end{figure}

\begin{table}[t!]
 \centering
 \footnotesize
 \renewcommand{\arraystretch}{0.95}
 \caption{$BVRIc$ band de-reddened flux densities of the CSPN's SED
 derived from the residual FEROS spectrum. The extinction free $V$-band magnitude ($m_{V}$)
 is \mbox{14.51~$\pm$~0.17}, where $F_{\lambda}$($m_{V}$ = 0) is
 3.631(--9) erg~s$^{-1}$~cm$^{-2}$~{\AA}$^{-1}$. \label{T:feros2}}
 \tablewidth{\textwidth}
\begin{tabular}{@{}ccc@{}}
\hline\hline
$\lambda_{c}$ ({\AA})&Band &$F_{\lambda}$ (erg s$^{-1}$ cm$^{-2}$ {\AA}$^{-1}$)\\
\hline
4378.1&	Johnson-$B$&	1.21(--14) $\pm$ 1.14(--15)\\
5466.1&	Johnson-$V$&	5.69(--15) $\pm$ 8.76(--16)\\
6358.0&	Cousins-$R$&	3.58(--15) $\pm$ 8.66(--16)\\
7829.2&	Cousins-$I$&	1.63(--15) $\pm$ 7.08(--16)\\
 \hline
 \end{tabular}
\end{table}

 First, we investigated the SED of the CSPN using the FEROS spectrum,
 which is the sum of the nebular emission lines and continuum and the
CSPN spectrum. For this end, we need to subtract the nebular continuum from the FEROS spectrum.
We used the {\sc Nebcont} code in the {\sc Dispo} package of
{\sc STARLINK} v.2015A\footnote{\url{http://starlink.eao.hawaii.edu/starlink}} to generate the nebular continuum. For the calculation,
we adopted the {\hb} flux of the entire nebula
\mbox{9.83(--12)\,erg~s$^{-1}$~cm$^{-2}$},
$n$(He$^{+}$/H$^{+}$) = 9.69(--2), {\te} = 8090\,K,
and {\Ne} = 22\,860\,cm$^{-3}$, which is the average among
{\Ne}({\siii}), {\Ne}({\cliii}), {\Ne}({\neiii}), and {\Ne}({\ariv}).

In Fig.~\ref{S:feros2} upper panel, we show the synthesized nebular
continuum. The discontinuity around 8200\,{\AA} indicates the Paschen
jump. After we had scaled the {\it de-reddened} FEROS spectrum up to
match with the
{\hb} line flux of the entire nebula, we manually removed
gas emission lines to the extent possible.
This gas emission line free FEROS spectrum
is presented in the same panel. In the lower panel, we show the
resultant spectrum generated by subtracting the synthesis nebular
continuum spectrum from the gas emission line free and flux density
scaled FEROS spectrum. Note that the residual
spectrum coincides with the spectrum of the CSPN. A spike
feature around 8200 {\AA} is from the residuals of Paschen and Bracket
continuum between the observed and the model. If we can subtract this
continuum around Paschen jump from the observed
spectrum, the spike feature will be gone. This spike feature does not affect
$Ic$-band magnitude measurement.
Using this residual spectrum, we measured flux densities for $BVRIc$
bands by taking filter transmission curves of each band, as summarized
in Table~\ref{T:feros2}.

\subsection{Synthesis of the CSPN's SED / Core-Mass / Distance \label{S:cspn2}}

\begin{figure}[t!]
   \includegraphics[width=\columnwidth,clip]{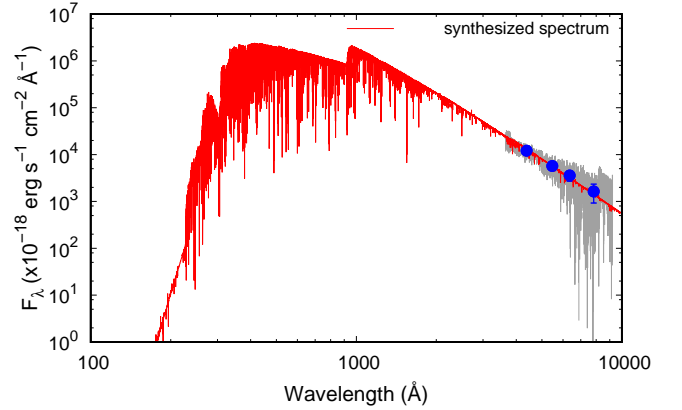}
 \caption{The spectrum of the CSPN synthesized using Tlusty (red
 line) in $T_{\rm eff}$ = 45\,000\,K and $\log\,g$ =
 5.25\,cm\,s$^{-2}$. The flux density is scaled down
 to the $F_{\lambda}$ = 5.69(--15)\,erg~s$^{-1}$~cm$^{-2}$
 {\AA}$^{-1}$at 5466.1\,{\AA}.
 The grey line and blue circles are the same as indicated in Fig~\ref{S:feros2}.
\label{S:tlusty}}
\end{figure}

\citet{Reindl:2014aa} performed spectral synthesis fitting of the
spectrum of the CSPN taken using Far Ultraviolet Spectroscopic
Explorer (\emph{FUSE}) in 2006, and they obtained $T_{\rm eff}$
= 55\,000\,K and $\log$\,$g$ =
\mbox{6.0~$\pm$~0.5\,cm\,s$^{-2}$}. However, in our Cloudy
model with this $T_{\rm eff}$ and the measured de-reddened $m_{V}$
of the CSPN (14.51, see \S \ref{S:cspn}) determining
the luminosity, we overproduced the fluxes of higher IP ions
such as {\neiii} and {\oiii} lines\footnote{
For example, when we adopt $T_{\rm eff}$ = 55\,000\,K and
distance $D$ = 2.5\,kpc, Cloudy model predicted that the respective
$I$([Ne\,{\sc iii}]\,3869\,{\AA}) and $I$([O\,{\sc iii}]\,5007\,{\AA})
are 134.6 (40.3 in our FEROS observation) and 303.4 (145.5), and
the predicted ionization boundary radius was 4.1$\arcsec$ (1.28$\arcsec$
measured from the \emph{HST}/WFPC2 {\hb} image, see \S
\ref{Snebucond}).
Maybe, if we set $D$ $>$ 8.0\,kpc, the $T_{\rm eff}$ = 55\,000 model
could explain the observed line fluxes. However, If we set $D$ to be
8\,kpc,  we will classify Hen3-1357 as a halo PN and estimate the core-mass of
the central star to be $>$0.53-0.60\,$M_{\odot}$.
}.

It might be because the nebula ionization
structure is not yet fully changed by the recent very fast post-AGB
evolution of the CSPN. Although we firmly believe the results of
\citet{Reindl:2014aa}, we needed to adopt a SED of the
CSPN with a lower $T_{\rm eff}$ to reproduce the overall observed nebular line fluxes.
For instance, we estimated $T_{\rm eff}$ to be
\mbox{50\,560~$\pm$~2710\,K}
using the {\oiii}/H$\beta$ line ratio and the formula
established among PNe in the Large Magellanic Cloud by \citet{Dopita:1991aa}.

Therefore, we utilized the non-local thermodynamic equilibrium (non-LTE)
stellar atmospheres modeling code Tlusty
\citep{Hubeny:1988aa}\footnote{
\url{http://nova.astro.umd.edu}}
to obtain SED of the CSPN for our Cloudy
model. Using Tlusty, we constructed
line-blanketed, plane-parallel, and hydrostatic stellar atmosphere,
where we considered the He/C/N/O/Ne/Si/P/S/Fe abundances. We run a
grid model to cover $T_{\rm eff}$ from 43\,000 to 53\,000\,K in a
constant 1000\,K steps. Here, we adopted the observed nebular
$\epsilon$(He), $\epsilon$(N(CEL)), $\epsilon$(O(CEL)), $\epsilon$(Ne), and
$\epsilon$(S). We adopted the expected \mbox{$\epsilon$(C(CEL)) = 7.98}
(see Table~\ref{T:abund}). As \citet{Reindl:2014aa} reported, there is
no significant difference between the nebular and stellar He/C/N/O/S
abundances. We adopted stellar \mbox{$\epsilon$(Si) = 7.52} and
\mbox{$\epsilon$(P) = 4.42} derived by \citet{Reindl:2014aa}.
From the nebular \mbox{$\langle$[S,Ar/H]$\rangle$
= --0.23}, we adopted \mbox{$\epsilon$(Fe) = 7.23}. We interpret that
99\,$\%$ of the Fe-atoms in the stellar atmosphere is eventually locked as dust grains in
the nebula. We set the microturbulent
velocity to 10\,{\kms} and the rotational velocity to 20\,{\kms}.

Based on \citet{Reindl:2014aa,Reindl:2017aa},
\citet{Parthasarathy:1993aa}, and \citet{Karakas:2010aa}, the core-mass of the CSPN ($m_{\ast}$)
is $\sim$0.53-0.6\,$M_{\sun}$. Referring to the theoretical post-AGB
evolution tracks presented in Fig.~4 of \citet{Reindl:2017aa}, we adopted
\mbox{$\log\,g$ = 5.25\,cm~s$^{-2}$}, and we adopted
the distance \mbox{$D$ = 2.5\,kpc} to obtain $m_{\ast}$
$\sim$0.53-0.6\,$M_{\sun}$. $D$ has been determined in the range
between 826\,pc \citep[see][reference therein]{Reindl:2017aa} and
5.85\,kpc \citep{Frew:2016aa}, so far. When we adopt \mbox{$D$ =
826\,pc}, we have to set a very small inner radius of the nebula
to reproduce the observed {\hb} flux by setting a very small inner
radius, we overproduced fluxes of higher IP lines and obtained
hotter dust temperatures, accordingly causing lower dust
continuum fluxes. If $D$ is 5.0\,kpc, the
situation would become better than the case of $D$ = 826\,{pc}, and then
we can reproduce the observed line fluxes. However,
we have to set $\log\,g$ $\sim$4.5\,cm s$^{-2}$ in order
to obtain the above $m_{\ast}$ range. And Hen3-1357 would be classified as a
halo PN not a thin disk PN.

We verified our adopted $D$ of 2.5\,kpc. 
Following \citet{Quireza:2007aa}, we can classify 
Hen3-1357 into a Type II or III PN based on the observed 
$\epsilon$(He) and N/O ratio. Hen3-1357 would be a thin disk 
population. \citet{Quireza:2007aa} reported that the average 
peculiar velocity relative to the Galaxy rotation 
($\Delta$$V$) is $\sim$23\,{\kms} for Type IIb and $\sim$70\,{\kms} for 
Type III and the average height from the Galactic plane ($|z|$) 
is $\sim$0.225\,kpc for Type IIa and $\sim$0.686\,kpc for Type III,
respectively. From the constraint on $|z|$, we obtained a range of 
$D$ toward Hen3-1357 between 1.07 and 3.27\,kpc. \citet{Maciel:2005aa}
calculated the rotation velocities at the nebula Galactocentric 
positions calculated for a Galaxy disk rotation curve 
based on four distance scales. Using their established Galaxy
rotation velocity based on the distance scale of
\citet{Cahn:1992aa}, \citet{van-de-Steene:1995aa}, and 
\citet{Zhang:1995aa}, equation 
(3) of \citet{Quireza:2007aa}, and our measured LSR radial velocity
12.29\,{\kms} (see \S \ref{S:flux}), we obtained a $D$ versus
$\Delta$$V$ plot. Using this plot and the constraint on $\Delta$$V$, we got 
another range of $D$ between 1.63 and 4.92\,kpc. 
Thus, we obtained \mbox{$D$ = 1.63-3.27\,kpc}, finally. 2.5\,kpc is the
middle value of this distance range.
From the above discussion, we adopted \mbox{$D$ = 2.5\,kpc} 
and the absolute $V$-band magnitude of the CSPN \mbox{$M_{V}$ = 2.555}.

Finally, we obtained the synthesized spectra using {\sc
SYNSPEC}\footnote{\url{http://nova.astro.umd.edu/Synspec49/synspec.html}} as displayed in
Fig.~\ref{S:tlusty}.

\subsection{Parameters of the nebular gas and dust grain \label{S:modelneb}}

\subsubsection{Nebular elemental abundances}

We adopted elemental abundances listed in Table~\ref{T:abund} as a first guess. We refined these abundances
to reproduce the observed emission line intensities. For the other
elements unseen in the FEROS and \emph{Spitzer}/IRS spectra, we referred
to the predicted values in the AGB nucleosynthesis model for
initially 1.5\,$M_{\sun}$ stars with \mbox{$Z$ = 0.008} by
\citet{Karakas:2010aa}. For the sake of consistency, we substituted the
transition probabilities and effective collision strengths of CELs by
the same values applied in our nebular abundance analysis.

In spite of non-detection in the \emph{Spitzer}/IRS spectrum,
our Cloudy model with the AGB nucleosynthesis predicted
$\epsilon$(Si) overestimated the
$[$Si\,{\sc iii}$]$\,34.82\,{\micron} line. This indicates
that most of the Si-atoms exist as amorphous silicate dust grains. Therefore,
we took care of the Si and Mg abundances as silicate grain components.
Assuming that the nebular [Mg,Si/H] is comparable to the [Mg/H] =
--1.69 measured in the PN IC4846 \citep{Hyung:2001aa}, we kept
$\epsilon$(Mg) = 5.86 and $\epsilon$(Si) = 5.84, respectively. As we
discussed later, IC4846 displays amorphous silicate features \citep[e.g.,][]{Stanghellini:2012aa} and very similar
elemental abundances to Hen3-1357.

\subsubsection{Nebula geometry/boundary condition/gas filling factor \label{Snebucond}}

We adopted spherical shell with a uniform hydrogen
density. We assumed the ionization boundary radius ($r_{\rm ib}$) of
 $\sim$1.3{\arcsec} using a plot of count versus size of the circular
 aperture generated by the archival \emph{HST}/Wide Field Planetary Camera 2
 (WFPC2) F487N ({\hb}) image taken on 1996 March 3 (Prop-ID: GO6039, PI: 
M.~Bobrowsky). 85\,{\%} of the total count is measured within 1.28{\arcsec}.
Although the exact size of the nebula in 2006 is unknown, slow nebula
shell expansion velocity suggests that the size of the nebula
is not largely different since 1996. Here, we measured twice expansion velocities (2$V_{\rm
exp}$) using equation (3) of \citet[e.g.,][]{Otsuka:2003aa,Otsuka:2009aa,Otsuka:2015aa} and 144 emission
lines as summarized in appendix Table~\ref{T:exp}. To calculate line
broadening by gas thermal motion, we adopted suitable {\te} for each
ion by referring to Table~\ref{T:diagno}. In Hen3-1357, 2$V_{\rm exp}$ did not
correlate with IP. We measured the average 2$V_{\rm
exp}$ = \mbox{14.8~$\pm$~0.5\,{\kms}} among the 39 H\,{\sc i}
lines, which is consistent with the mean expansion velocity
($V_{\rm exp}$) of 8.4~$\pm$~1.5\,{\kms} measured from the 17 lines by  \citet{Arkhipova:2013aa}.

Filling factor $f$ can be defined as the ratio
of an RMS density derived from a hydrogen
line flux, $T_{\rm e}$, and nebula radius to the $n_{\rm e}$(CELs)
\citep[see, e.g.,][]{Mallik:1988aa,Peimbert:2000aa}.
We calculated an RMS density of 10\,750\,cm$^{-3}$ from the {\hb} flux of the
entire nebula, $T_{\rm e}$ = 8\,060\,K, $r_{\rm ib}$ = 1.28{\arcsec}, and 
a constant $n_{\rm e}$/$n({\rm H^{+}})$ = 1.15. We estimated $f$
to be 0.47-0.62 using this RMS density and the observed $n_{\rm
e}$(CELs). Here, we set $f$ = 0.55 as a first guess and varied.

\subsubsection{Dust grains and size distribution}

We assumed spherical shaped silicate grain and adopted a
standard interstellar size distribution \citep[$n(a) \propto
a^{-3.5}$,][]{Mathis:1977aa} with radius $a$ 
= 0.01-0.50\,{\micron}. We selected the dielectric function 
table of astronomical silicate currently 
recommended by the webpage of \mbox{B.~Draine}
\footnote{\url{https://www.astro.princeton.edu/~draine/dust/dust.diel.html}}.

\subsection{Model result \label{S:resmodel}}

\begin{figure*}
\centering
 \includegraphics[width=0.8\textwidth]{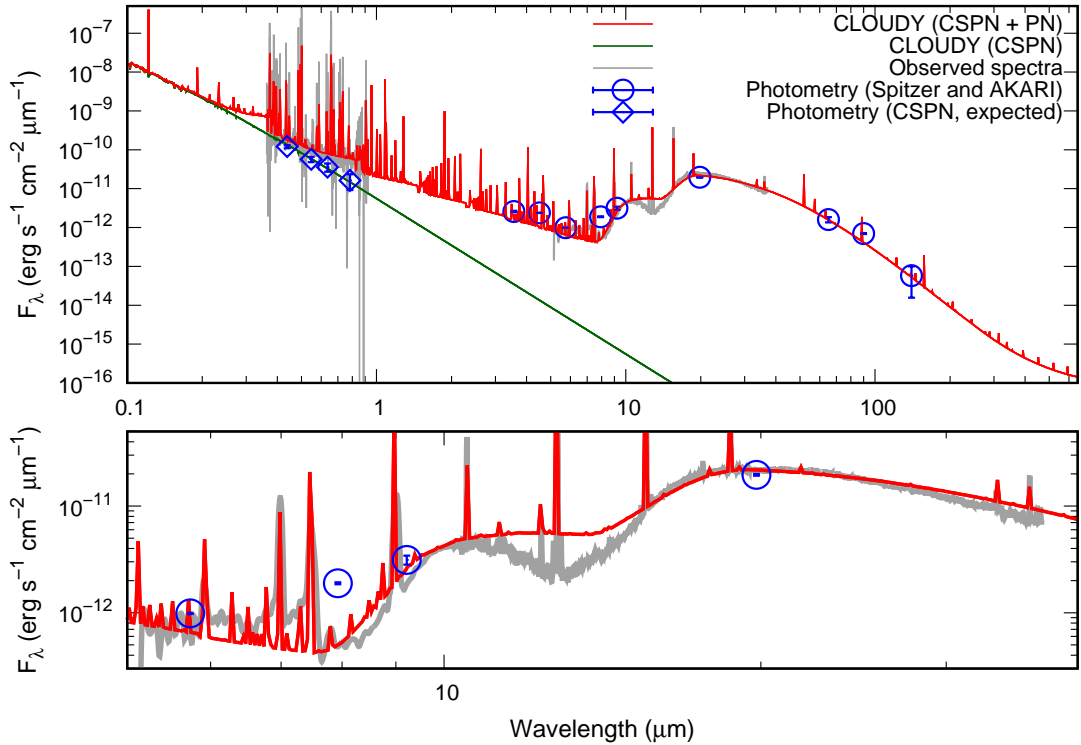}
\caption{
({\it upper panel}) Comparison between the Cloudy model and
 observational data of Hen3-1357. The blue diamonds indicate the
 observed $BVRIc$ band flux densities of the CSPN, which are the same
 values in listed in Table~\ref{T:feros2}. ({\it lower panel}) Closed-up
 plots in 5-40\,{\micron}. In both panels, we set the spectral
 resolution of the synthesized Cloudy spectrum to be a constant 600, corresponding to that of
 \emph{Spitzer}/IRS SH and LH spectra. \label{F:SED}}
\end{figure*}

\begin{table}[t!]
\centering
 \caption{
\label{T:model}
 The best-fit Cloudy model parameters of Hen3-1357.
}
 \footnotesize
 \renewcommand{\arraystretch}{0.95}
 \tablewidth{\columnwidth}
 \begin{tabular}{@{}ll@{}}
\hline\hline
{Parameters of the CSPN}      &\multicolumn1c{Value}\\
\hline
$L_{\ast}$ / $T_{\rm eff}$ / $\log\,g$ / $D$ &330\,$L_{\sun}$ / 45\,550\,K /
 5.25\,cm s$^{-2}$ / 2.5\,kpc\\
$M_{V}$ &2.555\\
$R_{\ast}$  &0.291\,$R_{\sun}$\\
$m_{\ast}$ &0.550\,$M_{\sun}$\\
\hline
{Parameters of the Nebula}      &\multicolumn1c{Value}\\
\hline
$\epsilon$(X)   &He:10.97, C:8.18, N:7.89, O:8.58,  Ne:8.20\\
                &Mg:5.86,  Si:5.84, S:6.74, Cl:4.73, Ar:6.25\\
                &Fe:5.23, Others: \citet{Karakas:2010aa}\\
Geometry        &Spherical symmetry\\
Shell size      &$r_{\rm in}$:0.44{\arcsec} (0.005\,pc), $r_{\rm out}$:2.77{\arcsec} (0.034\,pc)\\ 
Ionization boundary  &1.48{\arcsec} (0.018\,pc)\\
radius ($r_{\rm ib}$)\\
Filling factor ($f$) &0.58\\
$n_{\rm H}$     &11\,610\,cm$^{-3}$\\ 
$F$({\hb}) &9.84(--12)\,erg s$^{-1}$ cm$^{-2}$ (de-reddened)\\
$m_{g}$     &3.81(--2)\,$M_{\sun}$ \\
\hline
{Parameters of the Dust}      &\multicolumn1c{Value}\\
\hline
Grain size    &0.01-0.50\,{\micron}\\
$T_{d}$       &40-150\,K\\  
$m_{d}$       &1.98(--4)\,$M_{\sun}$\\ 
$m_{d}$/$m_{g}$ (DGR)&5.20(--3) \\
\hline
 \end{tabular}
\end{table}

    \begin{table}
     \centering
 \caption{\label{T:modelicf} Comparison of the ICFs between the
    observation and the Cloudy model.}
 \footnotesize
 \renewcommand{\arraystretch}{0.95}
 \tablewidth{\columnwidth}
\begin{tabular}{@{}lcc@{}}  
\hline\hline
Elem.	&ICF(obs)&	ICF(model)\\
\hline
He	&1.00	&1.03       \\
C(RL)	&1.48 $\pm$ 0.22&	1.30\\
N(RL)	&1.48 $\pm$ 0.22&	1.62\\
N(CEL)	&3.09 $\pm$ 0.17&	2.69\\
O(RL)	&2.45 $\pm$ 0.07&	2.23\\
 O(CEL)	&1.00	&1.04\\
 Ne	&1.00	&1.01\\
 S	&1.00	&1.00\\
 Cl	&1.01 $\pm$ 0.06&	1.01\\
 Ar	&1.00	&1.01\\
 Fe	&2.30 $\pm$ 0.14&	2.07\\
\hline
\end{tabular}
    \end{table}

  To find the best-fit model, we varied
  $T_{\rm eff}$, the inner radius of the nebula $r_{\rm in}$, $n_{\rm
  H}$, $\epsilon$(He/C/N/O/Ne/S/Cl/Ar/Fe), dust mass fraction, and $f$
  within a given range by using the optimize command available in Cloudy.

\citet{Garcia-Hernandez:2002aa} found that the distribution of 
 molecular hydrogen H$_{2}$ $v$=1-0 S(1)
 at 2.122\,{\micron} and $v$=2-1S (1) at 2.248\,{\micron} is
 quite homogeneous and extends well
 beyond the distribution of the H\,{\sc i} Br$\gamma$ line. This suggests
 that Hen3-1357 has large neutral regions.

Thus, we went to deep neutral gas regions in our model; 
we continued calculation until any of the model's predicted flux densities at
\emph{AKARI}/FIS 65/90/140\,{\micron} bands reached or exceeded
the relevant observed values. Cloudy model predicted $r_{\rm ib}$
= 1.48{\arcsec} where {\te} drops below 4000\,K. We stopped model
calculation at the outer radius ($r_{\rm out}$) of 3.4(--2) pc
(2.77{\arcsec}). The goodness of fit was determined by the reduced $\chi^{2}$
value calculated from the following observational constraints:
17 broadband fluxes, 5 broadband flux
densities, 104 gas emission fluxes, $r_{\rm ib}$, de-redden 
$F$({\hb}) of the entire nebula. Table\,\ref{T:model} summarizes
the parameters of the best-fit model, where the reduced $\chi^{2}$ is 33.5.

The SED of the best-fit model, in comparison with the observational
data is presented in Fig.~\ref{F:SED}. From the model
result, we confirmed that gas emission contribution to
\emph{Spitzer}/IRAC 8.0\,{\micron} and \emph{AKARI}/IRC
9.0/18\,{\micron} bands is
51.8\,\%, 19.1\,\%, and 3.9\,\%, respectively. Thus, the disagreement at
\emph{Spitzer}/IRAC 8.0\,{\micron} band between the observed photometry
and the predicted SED can be explained by considering
the gas emission contribution to the relevant band.

The observed and model predicted line fluxes, band fluxes, and band flux densities are summarized in appendix
Table~\ref{T:model2}. The intensity of the
O\,{\sc ii}\,4075\,{\AA} and 4651\,{\AA} is the sum of the multiplet
V10 and V1 O\,{\sc ii} lines, respectively. It is noteworthy that
we simultaneously reproduced both the observed RL/CEL N and O line fluxes.

The predicted ICF(X) by Cloudy listed in Table~\ref{T:modelicf} is
in excellent agreement with the ICF(X) derived in
\S \ref{S:elementICF}, indicating that our Cloudy model
succeeded to explain ionization nebula structure and the
ICF(X) based on IP is proper value.

As described in \S \ref{S:cspn2}, under the constraints to
the CSPN at $D$ = 2.5\,kpc, we need $T_{\rm eff}$ = 45\,550\,K and
$L_{\ast}$ = 330\,$L_{\odot}$ in order to explain the observed
quantities. With $D$, $L_{\ast}$, and $\log\,g$, 
we derived $m_{\ast}$ = 0.55\,$M_{\odot}$.

The gas mass ($m_{g}$) = 3.81(--2)\,$M_{\sun}$ is the sum of
the ionized and neutral gas masses. The ionized gas mass is 5.38(--3)\,$M_{\sun}$ and 
the remaining is the neutral gas mass. Our derived $m_{g}$ is close to
the ejected mass = 8.9(--2)\,$M_{\odot}$ in initially 1.5\,$M_{\odot}$
stars with $Z$ = 0.008 during the last thermal pulse AGB, predicted by
\citet{Karakas:2007aa}. We obtained the dust mass
($m_{d}$) of 1.98(--4)\,$M_{\sun}$.

It is of interest to know how far-IR data impact 
gas and dust mass estimates in our model. When we stopped
model calculation at $r_{\rm ib}$, we obtained 
$m_{g}$ = 4.61(--3)\,$M_{\sun}$ and $m_{d}$ = 3.79(--5)\,$M_{\sun}$,
respectively. This model did not well fit any \emph{AKARI} far-IR
fluxes. To fit the observed far-IR data, we need a larger $r_{\rm
out}$. With the \emph{AKARI} far-IR data, we obviously obtained much
greater $m_{g}$ and $m_{d}$. About 80\,$\%$ of the total dust mass is
from warm-cold dust components beyond ionization front. From the model
result, we confirmed that the gas emission contribution to
\emph{AKARI} 65/90/140\,{\micron} bands is 
1.4\,\%, 1.08\,\%, and 2.58\,\%, respectively.
\emph{AKARI} far-IR data would be thermal emission from warm-cold dust.

 \citet{Cox:2011aa} derived an upper limit of the sum of $m_{d}$ and
 $m_{g}$ = 0.16\,$M_{\sun}$ within a 3\,pc radius using
 the \emph{AKARI}/90\,{\micron} and a
 constant dust-to-gas mass ratio (DGR) = 6.25(--3)
 for O-rich dust, although the measured dust temperature ($T_{d}$) is unknown.
 Using their results, we calculated an upper limit $m_{g}$ =
 0.159\,$M_{\sun}$ and $m_{d}$ = 9.94(--4)\,$M_{\sun}$, respectively.

\citet{Umana:2008aa} derived the total ionized mass of
$\sim$5.7(--2)\,$M_{\sun}$ using the radio data in 2002,
assuming $D$ = 5.6\,kpc, inner/outer radii =
0.65{\arcsec}/1.3{\arcsec} shallow shell geometry, and $f$ = 1.0. Using
the \emph{IRAS} data, 
they derived \mbox{$T_{d}$ = 137~$\pm$~2\,K}, and \mbox{$m_{d}$ =
2(--4)\,$M_{\sun}$}
using the 60\,{\micron} flux density in the case of
silicate. Based on the results and assumptions of \citet{Umana:2008aa},
ionized gas and co-existing dust would be $\sim$6.6(--3)\,$M_{\sun}$ and
$\sim$4.0(--5)\,$M_{\sun}$, respectively if we adopt $D$ = 2.5\,kpc and $f$ =
0.58. These estimated values are consistent with our derived $m_{g}$
and $m_{d}$ when we stopped the model at $r_{\rm ib}$.
On the dust, \citet{Cox:2011aa} found
that the \emph{AKARI} far-IR flux densities are by a factor two
 lower than predicted from the \emph{IRAS} data. They interpreted that
 the far-IR variability in its infrared flux might occur due to recent
 mass-loss event(s) or evolution of the CSPN. Following the report of 
\citet{Cox:2011aa} and the dust mass $\sim$4.0(--5)\,$M_{\sun}$
co-existing with the ionized gas by the \emph{IRAS} data in 1980s, we estimate
the dust mass to be $\lesssim$~2(--5)\,$M_{\sun}$ in 2006-2007, which is comparable to 
our derived dust mass of 3.79(--5)\,$M_{\sun}$ within the ionized gas.

\section{Discussion}

\begin{figure}
\centering
\includegraphics[width=\columnwidth,clip]{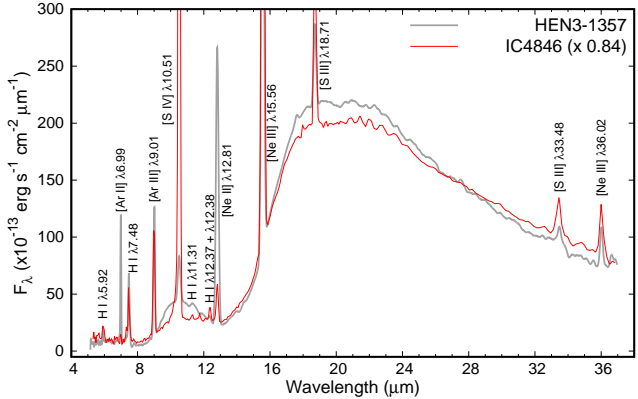}
\caption{
 The \emph{Spitzer}/IRS spectra of IC4846 and Hen3-1357.
 The spectral resolution of Hen3-1357 is down to match with that of IC4846.
 For IC4846, we scaled the flux density up
 to match with \emph{AKARI}/IRC 9.0/18\,{\micron} bands \citep[0.1311 and
 2.038 Jy, respectively,][]{Yamamura:2010aa}. For
 demonstration, the flux density of this scaled spectrum is further
 scaled to match with the IRS spectrum of Hen3-1357 by a constant factor of
 0.84. See text for details.
 \label{F:spt2}}
\end{figure}

 \begin{table}[t!]
  \begin{center}
 \footnotesize
\renewcommand{\arraystretch}{0.95}
 \tablewidth{\columnwidth}
\caption{Comparisons with the $\epsilon$(X) of IC4846 and
  the average $\epsilon$(X) value among Galactic amorphous silicate rich
   PNe. The $\epsilon$(X) of Hen3-1357 is the result by using the ICFs,
   except for the CEL C, which is an expected value.\label{T:comp}}
 \begin{tabular}{@{}l@{\hspace{1pt}}r@{\hspace{3pt}}r@{\hspace{3pt}}r
  @{\hspace{3pt}}r@{\hspace{3pt}}r@{\hspace{3pt}}r@{\hspace{3pt}}r@{\hspace{5pt}}r@{}}
 \hline\hline
Elem. &\multicolumn{1}{c}{OD PN}      &\multicolumn{6}{c}{IC4846}   &\multicolumn{1}{r}{Hen}\\
                  \cline{3-8}     
(X)   &\multicolumn{1}{c}{Ave.}       &\multicolumn{1}{c}{(a)} & \multicolumn{1}{c}{(b)}
	 &\multicolumn{1}{c}{(c)}
	     &\multicolumn{1}{c}{(d)} & \multicolumn{1}{c}{(e)} & \multicolumn{1}{c}{Ave.} &\multicolumn{1}{r}{3-1357}  \\ 
\hline
He     &11.02 &10.98 & 10.96 & 10.90 & 11.01 & \nodata & 10.96 & 10.99 \\ 
C(RL)  &\nodata &7.74 & 8.37 & 8.43 &\nodata  & \nodata & 8.27 & 8.16 \\ 
C(CEL) &\nodata &7.68 & 8.45 & 8.16 &\nodata  & 7.95 & 8.15 & 7.98 \\ 
N(RL)  &\nodata &\nodata & \nodata & 8.10 &\nodata  & \nodata & 8.10 & 8.01 \\ 
N(CEL) &7.78 &7.89 & 7.81 & 8.09 & 7.69 & \nodata & 7.90 & 8.05 \\ 
O(RL)  &\nodata &\nodata & 8.97 & 8.78 &  & \nodata & 8.89 & 8.84 \\ 
O(CEL) &8.42& 8.60 & 8.51 & 8.59 & 8.60 & 8.50 & 8.56 & 8.66 \\ 
Ne     &7.78& 7.90 & 7.83 & 7.77 & 7.99 & \nodata & 7.88 & 8.19 \\ 
Mg     &\nodata &5.86 &\nodata  &\nodata  &\nodata  &\nodata  & 5.86 & \nodata \\ 
S      &6.50& 6.95 & 6.63 & 7.01 & 6.73 & \nodata & 6.86 & 6.83 \\ 
Cl     &6.15& 5.11 & \nodata & 5.34 & 6.14 & \nodata & 5.76 & 5.08 \\ 
Ar     &6.03& 6.18 & 5.96 & 6.02 & 6.13 & \nodata & 6.08 & 6.37 \\ 
Fe     &\nodata &\nodata & \nodata & \nodata &\nodata  & 5.21 & 5.21 & 5.22 \\ 
\hline
\end{tabular}
  \end{center}
  
  \tablerefs{
The average abundance of Galactic
  amorphous silicate rich PNe
  in the Galaxy (OD PN Ave) in the second column is taken from
  \citet{Garcia-Hernandez:2014aa}. On elemental abundances of the PN
  IC4846 in the third - seventh columns - (a) \citet{Hyung:2001aa},
  (b) \citet{Wesson:2005aa}, (c) \citet{Wang:2007aa}, (d)
  \citet{Garcia-Hernandez:2014aa}, and (e)
   \citet{Delgado-Inglada:2014aa}. The eighth column is the average
  among the measurements by (a)-(e).}
 \end{table}

It is necessary to verify whether the gas and dust chemistry in
Hen3-1357 is consistent with other O-rich gas and dust Galactic
PNe. To compare with such PNe
is an important step to understand the evolution of Hen3-1357.

\citet{Garcia-Hernandez:2014aa} investigated relations among dust
features, elemental abundances, and evolution of the progenitors.
In the second column of Table~\ref{T:comp}, we list the average $\epsilon$(X)
among their amorphous silicate PNe. They found that $\epsilon$(He) and
N/O ratio in these amorphous silicate containing PNe are in agreement
with the AGB nucleosynthesis model predictions
for initially $\sim$1.0\,$M_{\sun}$ stars with $Z$ = 0.008.
\citet{Garcia-Hernandez:2014aa} suggested that the higher
Ne/O ratios in O-rich dust PNe relative to the AGB models may reflect
the effect of PMZ. The observed $\epsilon$(He) and the CEL
N/O ratio of \mbox{0.24~$\pm$~0.02} in Hen3-1357 coincide
with the average values in their amorphous silicate
PN sample. As discussed in \S \ref{S:AGB}, our predicted
progenitor mass, initial metallicity,
and interpretation for the Ne overabundance in Hen3-1357 follow their results.

We can now understand relations among dust features,
nebular abundances, and the progenitor stars' evolution.
Moreover, we know that the nebula morphology is connected to the central
star's evolution. Using the \emph{HST}/WFPC images
as a guide, we tried to find objects showing
similar nebula shape, dust features, and elemental abundance pattern
to Hen3-1357. As far as our best knowledge, a point-symmetric
PN IC4846 \citep[e.g.,][]{Miranda:2001aa} is very similar
to Hen3-1357.

IC4846 clearly shows amorphous silicate features as reported by
\citet{Stanghellini:2012aa}. We reduced the BCD of IC4846
(obs AORKEY: 25839616, PI: L.~Stanghellini) by the same process applied
for Hen3-1357.  In Fig.~\ref{F:spt2}, we display the \emph{Spitzer}/IRS
spectra of IC4846 and Hen3-1357. The dust features
seen in both PNe are very similar except for the different strengths
of the 9 and 18\,$\mu$m emission bumps, which might reflect the
difference in the grain composition. 
For IC4846, \citet{Stasinska:1999aa} derived a single \mbox{$T_{d}$ =
107\,K} and \mbox{DGR = 1.2(--3)} based on the \emph{IRAS} four
band fluxes using a modified blackbody function.
\citet{Tajitsu:1998aa} and derived a single \mbox{$T_{d}$ = 168\,K} using
the \emph{IRAS} data. \citet{Zhang:1991aa} derived a single \mbox{$T_{d}$ =
152\,K} and \mbox{$T_{\rm eff}$ = 47\,600\,K} by fitting SED from
\emph{IUE} to \emph{IRAS} data.

In the third to seventh columns of Table~\ref{T:comp}, we compile
nebular abundances of IC4846 measured by prior works. The eighth column
gives the average value. Obviously, the abundances in both IC4846 and
Hen3-1357 are in
excellent agreement even in the RL $\epsilon$(C,N,O) and the
Fe-depletion. So far, the $\epsilon$(Mg) and $\epsilon$(Fe) measurements
have been performed only by \citet{Hyung:2001aa} using the \emph{IUE}
UV-spectrum and only by \citet{Delgado-Inglada:2014aa} using the optical
spectra, respectively. The largely depleted [Mg/H] = --1.69 in IC4846
might indicate that most of the Mg-atoms are captured by silicate
grains. We assumed the similar situation to Hen3-1357 in our Cloudy model.

\citet{Hyung:2001aa} succeeded to reproduce UV-optical gas emission line fluxes in
photoionization model of IC4846 by setting the CSPN's radius 0.425\,$R_{\sun}$,
\mbox{$T_{\rm eff}$ = 70\,000\,K}, \mbox{$\log\,g$ = 4.6\,cm\,s$^{-2}$},
and \mbox{$D$ = 7\,kpc}, which give \mbox{$L_{\ast}$ =
3900\,$L_{\sun}$}.  With comparison with post-AGB evolutionary tracks,
they estimated $m_{\ast}$ $\sim$0.57\,$M_{\sun}$.

From above comparisons, we can conclude that Hen3-1357 is an ordinary
amorphous silicate rich and O-rich gas PN. Among amorphous silicate
rich PNe in the Milky Way, IC4846 is very similar to Hen3-1357.
Both PNe have evolved from similar progenitor mass stars with $Z$ =
0.008. However, the rapid evolution of the central star
of Hen3-357 still remains a puzzle.

\section{Summary}

We performed a detailed chemical abundance analysis and
constructed the photoionization model of Hen3-1357 to
characterize the PN and obtain a coherent picture of the
dusty nebula and CSPN in 2006 based on optical to far-IR data.

We calculated the abundances of the nine elements. The RL C/O ratio indicates that
Hen3-1357 is an O-rich PN, supported by the detection
of the broad 9/18\,$\mu$m amorphous silicate bands in the \emph{Spitzer}/IRS
 spectrum. The ADF(O$^{2+}$) is less than a typical value measured in PNe.
 The observed elemental abundances can be explained by AGB
 nucleosynthesis models of \cite{Karakas:2010aa} for initially 1-1.5\,$M_{\odot}$ stars with $Z$
 = 0.008. The Ne overabundance might be due to the enhancement of
  $^{22}$Ne isotope in the He-rich intershell.
 
 We did not find significant variation of nebular line intensities
 between 2006 and 2011, suggesting that
 nebular ionization state and elemental abundances are most
 likely in a steady state during the same period, while the central star is rapidly evolving.

 By incorporating the spectrum of the CSPN synthesized by Tlusty
 as the ionization/heating source of the PN with Cloudy modeling,
 we succeeded to explain the observed SED and derive
 the gas and dust masses, dust-to-gas mass ratio, and core-mass of the
 CSPN. About 80\,\% of the total dust mass is from the warm-cold dust
 components beyond ionization front.

 Through comparison with other Galactic PNe, we found that Hen3-1357 is
 an ordinary amorphous silicate rich and O-rich gas PN. IC4846 shows
 many similarities in properties of the PN to Hen3-1357.

 Although we derived physical properties of the nebula and also
 provided the range of the progenitor mass,
 the rapid evolution from post-AGB B1 supergiant
 in 1971 to a young PN in a matter of 21 years is not yet understood.
If the central star has experienced LTP then it should be
 H-poor, He and C-rich in its present hot post-AGB stage soon after
 the LTP. However, the nebular and stellar chemical compositions calculated by us and
 \citet{Reindl:2014aa,Reindl:2017aa} are nearly solar,
 not at all similar to those of LTP PNe.
If the central star has now started returning towards the
AGB phase, then very soon it will go through A, F, and G spectral types before
it appears as a born-again AGB star. If so, it may show abundances
similar to that of LTP PN in future.
We need to monitor the
central star's $T_{\rm eff}$, $\log\,g$, and chemical composition in
order to confirm whether it is
 evolving back towards the AGB stage. If Hen3-1357 is a binary,
 rapid evolution might be explained. For that end, monitoring
 of radial velocity using stellar absorption profiles in UV wavelength would be necessary.
 Moreover, comparisons with
 other Galactic amorphous silicate rich and O-rich gas PNe such as IC4846 
can help us to understand the evolution of Hen3-1357. Thus, further
observations of both the nebula and the central star are required for
 further understanding this PN.

\section*{Acknowledgments}
We are grateful to the anonymous referee for a careful reading and
valuable suggestions.
MO thanks Prof.~Ivan Hubeny for useful suggestions on Tlusty
modeling. MO was supported by the research fund 104-2811-M-001-138 and
104-2112-M-001-041-MY3 from the Ministry of Science and Technology
(MOST), R.O.C. This work was partly based on archival data obtained with
the \emph{Spitzer} Space Telescope, which is operated by the Jet 
Propulsion Laboratory, California Institute of Technology under a 
contract with NASA. This research is in part based on observations with 
AKARI, a JAXA project with the participation of ESA. 
Support for this work was provided by an award 
issued by JPL/Caltech. Some of the data used in this paper were
obtained from the Mikulski Archive for Space Telescopes (MAST).
STScI is operated by the Association of Universities for Research
in Astronomy, Inc., under NASA contract NAS5-26555. Support for MAST
for non-HST data is provided by the NASA Office of Space Science via
grant NNX09AF08G and by other grants and contracts.
A portion of this work was based on 
the use of the ASIAA clustering computing system.

\software{
IRAF (v.2.16),
SMART \citep[v.8.2.9:][]{Higdon:2004aa},
IRSCLEAN (v.2.1.1),
MOPEX,
STARLINK (v.2015A),
CLOUDY \citep[v13.03:][]{Ferland:2013aa},
TLUSTY \citep{Hubeny:1988aa}
}

\bibliographystyle{aasjournal}

\clearpage
\appendix

\setcounter{table}{0}
 \begin{deluxetable}{@{}clcrrr@{}}
\tablecaption{The identified atomic emission lines in the FEROS
 spectrum. The first column is the wavelength at the observation. The
 third column is the wavelength at rest in laboratory.
 \label{T:feros}}
  \renewcommand{\arraystretch}{0.95}
 \tablecolumns{6}
\renewcommand{\thetable}{A\arabic{table}}
\tabletypesize{\footnotesize}
 \tablewidth{\textwidth}
\tablehead{
$\lambda_{\rm obs.}$ 
&\multicolumn1c{Line}
&$\lambda_{\rm lab.}$
&\multicolumn1c{$f$($\lambda$)}
&\multicolumn1c{$I$($\lambda$)}
&\multicolumn1c{$\delta$$I$($\lambda$)}\\      
({\AA})&
&({\AA})&
 &\multicolumn2c{($I$({\hb}) = 100)}
 }
 \startdata
3697.33 & H\,{\sc i} (B17) & 3697.15 & 0.328 & 1.923 & 0.107 \\ 
3704.01 & H\,{\sc i} (B16) & 3703.85 & 0.327 & 2.028 & 0.122 \\ 
3705.16 & He\,{\sc i}        & 3704.98 & 0.327 & 0.965 & 0.087 \\ 
3712.12 & H\,{\sc i} (B15) & 3711.97 & 0.325 & 2.384 & 0.119 \\ 
3722.02 & H\,{\sc i} (B14) & 3721.94 & 0.323 & 4.156 & 0.155 \\ 
3723.81 & Fe\,{\sc ii}]      & 3723.92 & 0.323 & 0.474 & 0.058 \\ 
3726.20 & [O\,{\sc ii}]     & 3726.03 & 0.322 & 101.210 & 2.625 \\ 
3728.96 & [O\,{\sc ii}]     & 3728.81 & 0.322 & 37.647 & 0.979 \\ 
3734.52 & H\,{\sc i} (B13) & 3734.37 & 0.321 & 3.026 & 0.097 \\ 
3750.31 & H\,{\sc i} (B12) & 3750.15 & 0.317 & 4.220 & 0.124 \\ 
3770.79 & H\,{\sc i} (B11) & 3770.63 & 0.313 & 4.353 & 0.115 \\ 
3798.05 & H\,{\sc i} (B10) & 3797.90 & 0.307 & 5.330 & 0.133 \\ 
3819.79 & He\,{\sc i}        & 3819.60 & 0.302 & 1.142 & 0.033 \\ 
3833.75 & He\,{\sc i}        & 3833.55 & 0.299 & 0.072 & 0.010 \\ 
3835.54 & H\,{\sc i} (B9) & 3835.38 & 0.299 & 7.597 & 0.183 \\ 
3867.69 & He\,{\sc i}        & 3867.47 & 0.291 & 0.147 & 0.008 \\ 
3868.92 & [Ne\,{\sc iii}]    & 3869.06 & 0.291 & 40.325 & 0.946 \\ 
3871.95 & He\,{\sc i}        & 3871.79 & 0.290 & 0.078 & 0.006 \\ 
3889.06 & H\,{\sc i} (B8) & 3889.05 & 0.286 & 15.997 & 0.472 \\ 
3926.71 & He\,{\sc i}        & 3926.54 & 0.277 & 0.124 & 0.005 \\ 
3964.90 & He\,{\sc i}        & 3964.73 & 0.267 & 0.540 & 0.012 \\ 
3967.63 & [Ne\,{\sc iii}]    & 3967.79 & 0.267 & 10.000 & 0.216 \\ 
3970.23 & H\,{\sc i} (B7) & 3970.07 & 0.266 & 16.026 & 0.341 \\ 
4009.42 & He\,{\sc i}        & 4009.26 & 0.256 & 0.152 & 0.006 \\ 
4026.37 & He\,{\sc i}        & 4026.20 & 0.251 & 1.532 & 0.031 \\ 
4068.77 & [S\,{\sc ii}]     & 4068.60 & 0.239 & 4.471 & 0.086 \\ 
4069.74 & O\,{\sc ii}       & 4069.62 & 0.239 & 0.027 & 0.003 \\ 
4070.04 & O\,{\sc ii}       & 4069.88 & 0.239 & 0.038 & 0.005 \\ 
4072.32 & O\,{\sc ii}       & 4072.15 & 0.238 & 0.057 & 0.003 \\ 
4076.02 & O\,{\sc ii}       & 4075.86 & 0.237 & 0.076 & 0.005 \\ 
4076.52 & [S\,{\sc ii}]     & 4076.35 & 0.237 & 1.508 & 0.029 \\ 
4097.45 & N\,{\sc iii}      & 4097.35 & 0.231 & 0.023 & 0.003 \\ 
4101.90 & H\,{\sc i} (B6, H$\delta$)& 4101.73 & 0.230 & 21.516 & 0.395 \\ 
4103.19 & O\,{\sc ii}       & 4103.00 & 0.229 & 0.030 & 0.004 \\ 
4103.81 & N\,{\sc iii}      & 4103.39 & 0.229 & 0.028 & 0.004 \\ 
4105.13 & O\,{\sc ii}       & 4104.99 & 0.229 & 0.040 & 0.004 \\ 
4119.45 & O\,{\sc ii}       & 4119.22 & 0.224 & 0.019 & 0.004 \\ 
4121.00 & He\,{\sc i}        & 4120.81 & 0.224 & 0.177 & 0.005 \\ 
4143.93 & He\,{\sc i}        & 4143.76 & 0.217 & 0.229 & 0.005 \\ 
4153.44 & O\,{\sc ii}       & 4153.30 & 0.214 & 0.031 & 0.002 \\ 
4267.35 & C\,{\sc ii}       & 4267.18 & 0.180 & 0.103 & 0.005 \\ 
4276.00 & O\,{\sc ii}       & 4275.99 & 0.177 & 0.032 & 0.005 \\ 
4340.64 & H\,{\sc i} (B5, H$\gamma$)& 4340.46 & 0.157 & 46.052 & 0.579 \\ 
4349.64 & O\,{\sc ii}       & 4349.43 & 0.154 & 0.033 & 0.002 \\ 
4363.38 & [O\,{\sc iii}]    & 4363.21 & 0.149 & 2.461 & 0.030 \\ 
4367.08 & O\,{\sc ii}       & 4366.90 & 0.148 & 0.034 & 0.003 \\ 
4368.41 & O\,{\sc i}        & 4368.24 & 0.148 & 0.036 & 0.002 \\ 
4388.11 & He\,{\sc i}        & 4387.93 & 0.142 & 0.439 & 0.006 \\ 
4437.76 & He\,{\sc i}        & 4437.55 & 0.126 & 0.066 & 0.004 \\ 
4471.68 & He\,{\sc i}        & 4471.47 & 0.115 & 4.628 & 0.048 \\ 
4591.14 & N\,{\sc ii}       & 4590.85 & 0.078 & 0.036 & 0.005 \\ 
4630.71 & N\,{\sc ii}       & 4630.54 & 0.066 & 0.020 & 0.003 \\ 
4639.02 & O\,{\sc ii}       & 4638.86 & 0.064 & 0.034 & 0.004 \\ 
4642.01 & O\,{\sc ii}       & 4641.81 & 0.063 & 0.064 & 0.002 \\ 
4649.33 & O\,{\sc ii}       & 4649.13 & 0.061 & 0.103 & 0.003 \\ 
4651.01 & O\,{\sc ii}       & 4650.84 & 0.060 & 0.034 & 0.002 \\ 
4658.33 & [Fe\,{\sc iii}]    & 4658.05 & 0.058 & 0.111 & 0.002 \\ 
4661.83 & O\,{\sc ii}        & 4661.63 & 0.057 & 0.048 & 0.002 \\ 
4676.48 & O\,{\sc ii}       & 4676.23 & 0.053 & 0.032 & 0.005 \\ 
4701.86 & [Fe\,{\sc iii}]    & 4701.53 & 0.045 & 0.046 & 0.003 \\ 
4711.59 & [Ar\,{\sc iv}]     & 4711.37 & 0.042 & 0.034 & 0.003 \\ 
4713.38 & He\,{\sc i}        & 4713.22 & 0.042 & 0.668 & 0.007 \\ 
4725.73 & [Ne\,{\sc iv}]?     & 4725.64 & 0.038 & 0.011 & 0.002 \\ 
4734.06 & [Fe\,{\sc iii}]    & 4733.91 & 0.036 & 0.026 & 0.003 \\ 
4740.41 & [Ar\,{\sc iv}]     & 4740.16 & 0.034 & 0.070 & 0.002 \\ 
4754.94 & [Fe\,{\sc iii}]    & 4754.69 & 0.030 & 0.029 & 0.004 \\ 
4861.52 & H\,{\sc i} (B4, H$\beta$)        & 4861.33 & 0.000 & 100.000 & 0.112 \\ 
4881.20 & [Fe\,{\sc iii}]    & 4881.00 & -0.005 & 0.045 & 0.004 \\ 
4891.21 & O\,{\sc ii}? & 4890.86 & -0.008 & 0.032 & 0.008 \\ 
4922.13 & He\,{\sc i}        & 4921.93 & -0.016 & 1.214 & 0.004 \\ 
4924.78 & [Fe\,{\sc iii}]    & 4924.54 & -0.017 & 0.027 & 0.002 \\ 
4931.45 & [O\,{\sc iii}]    & 4931.23 & -0.019 & 0.055 & 0.003 \\ 
4959.13 & [O\,{\sc iii}]    & 4958.91 & -0.026 & 145.519 & 0.380 \\ 
4987.58 & [Fe\,{\sc iii}]    & 4987.21 & -0.033 & 0.016 & 0.003 \\ 
4996.95 & O\,{\sc ii}       & 4996.98 & -0.035 & 0.045 & 0.004 \\ 
5015.88 & He\,{\sc i}        & 5015.68 & -0.040 & 2.161 & 0.012 \\ 
5047.95 & He\,{\sc i}        & 5047.74 & -0.048 & 0.168 & 0.003 \\ 
5146.79 & [Fe\,{\sc iii}]    & 5146.45 & -0.071 & 0.027 & 0.004 \\ 
5159.01 & [Fe\,{\sc ii}]     & 5158.78 & -0.074 & 0.017 & 0.002 \\ 
5191.94 & [Ar\,{\sc iii}]    & 5191.82 & -0.081 & 0.064 & 0.003 \\ 
5198.13 & [N\,{\sc i}]       & 5197.90 & -0.082 & 0.363 & 0.004 \\ 
5200.49 & [N\,{\sc i}]      & 5200.26 & -0.083 & 0.228 & 0.004 \\ 
5270.74 & [Fe\,{\sc iii}]    & 5270.40 & -0.098 & 0.059 & 0.002 \\ 
5517.91 & [Cl\,{\sc iii}]    & 5517.72 & -0.145 & 0.117 & 0.003 \\ 
5538.06 & [Cl\,{\sc iii}]    & 5537.89 & -0.149 & 0.240 & 0.005 \\ 
5577.64 & [O\,{\sc i}]      & 5577.34 & -0.156 & 0.219 & 0.004 \\ 
5679.81 & N\,{\sc ii}        & 5679.56 & -0.173 & 0.017 & 0.002 \\ 
5754.83 & [N\,{\sc ii}]     & 5754.64 & -0.185 & 2.578 & 0.038 \\ 
5791.58 & C\,{\sc ii}        & 5791.69 & -0.190 & 0.021 & 0.002 \\ 
5875.88 & He\,{\sc i}        & 5875.60 & -0.203 & 14.666 & 0.262 \\ 
5958.74 & O\,{\sc i}         & 5958.54 & -0.215 & 0.026 & 0.004 \\ 
6300.55 & [O\,{\sc i}]      & 6300.30 & -0.263 & 16.129 & 0.339 \\ 
6312.32 & [S\,{\sc iii}]    & 6312.10 & -0.264 & 1.031 & 0.022 \\ 
6364.04 & [O\,{\sc i}]      & 6363.78 & -0.271 & 5.095 & 0.111 \\ 
6527.42 & [N\,{\sc ii}]     & 6527.24 & -0.293 & 0.019 & 0.003 \\ 
6548.32 & [N\,{\sc ii}]     & 6548.04 & -0.296 & 40.956 & 0.967 \\ 
6578.31 & C\,{\sc ii}       & 6578.05 & -0.300 & 0.050 & 0.004 \\ 
6583.70 & [N\,{\sc ii}]     & 6583.46 & -0.300 & 121.454 & 2.915 \\ 
6678.42 & He\,{\sc i}        & 6678.15 & -0.313 & 3.972 & 0.099 \\ 
6716.73 & [S\,{\sc ii}]      & 6716.44 & -0.318 & 6.066 & 0.154 \\ 
6721.76 & O\,{\sc ii}       & 6721.39 & -0.319 & 0.009 & 0.002 \\ 
6731.10 & [S\,{\sc ii}]     & 6730.81 & -0.320 & 12.471 & 0.319 \\ 
7002.42 & O \,{\sc i}         & 7002.12 & -0.356 & 0.051 & 0.004 \\ 
7062.57 & He\,{\sc i}        & 7062.28 & -0.364 & 0.019 & 0.003 \\ 
7065.50 & He\,{\sc i}        & 7065.18 & -0.364 & 7.846 & 0.247 \\ 
7136.06 & [Ar\,{\sc iii}]    & 7135.80 & -0.374 & 10.767 & 0.322 \\ 
7155.53 & [Fe\,{\sc ii}]     & 7155.16 & -0.376 & 0.034 & 0.003 \\ 
7160.85 & He\,{\sc i}        & 7160.61 & -0.377 & 0.027 & 0.003 \\ 
7254.70 & O \,{\sc i}        & 7254.45 & -0.390 & 0.064 & 0.004 \\ 
7281.64 & He\,{\sc i}        & 7281.35 & -0.393 & 0.726 & 0.023 \\ 
7298.33 & He\,{\sc i}        & 7298.04 & -0.395 & 0.035 & 0.003 \\ 
7319.33 & [O\,{\sc ii}]     & 7318.92 & -0.398 & 3.998 & 0.130 \\ 
7320.40 & [O\,{\sc ii}]     & 7319.99 & -0.398 & 13.487 & 0.430 \\ 
7329.96 & [O\,{\sc ii}]     & 7329.66 & -0.400 & 7.476 & 0.239 \\ 
7331.04 & [O\,{\sc ii}]     & 7330.73 & -0.400 & 6.987 & 0.224 \\ 
7378.30 & [Ni\,{\sc ii}]     & 7377.83 & -0.406 & 0.030 & 0.003 \\ 
7500.18 & He\,{\sc i}        & 7499.85 & -0.422 & 0.036 & 0.004 \\ 
7751.42 & [Ar\,{\sc iii}]    & 7751.10 & -0.455 & 2.630 & 0.096 \\ 
7816.45 & He\,{\sc i}        & 7816.14 & -0.464 & 0.077 & 0.003 \\ 
8245.95 & H\,{\sc i} (P42) & 8245.64 & -0.516 & 0.033 & 0.004 \\ 
8248.09 & H\,{\sc i} (P41) & 8247.73 & -0.516 & 0.027 & 0.003 \\ 
8250.17 & H\,{\sc i} (P40) & 8249.97 & -0.517 & 0.028 & 0.001 \\ 
8252.65 & H\,{\sc i} (P39) & 8252.40 & -0.517 & 0.033 & 0.002 \\ 
8255.36 & H\,{\sc i} (P38) & 8255.02 & -0.517 & 0.027 & 0.002 \\ 
8258.21 & H\,{\sc i} (P37) & 8257.85 & -0.517 & 0.041 & 0.002 \\ 
8261.20 & H\,{\sc i} (P36) & 8260.93 & -0.518 & 0.042 & 0.002 \\ 
8264.69 & He\,{\sc i}        & 8264.62 & -0.518 & 0.059 & 0.003 \\ 
8268.31 & H\,{\sc i} (P34) & 8267.94 & -0.519 & 0.055 & 0.003 \\ 
8272.22 & H\,{\sc i} (P33) & 8271.93 & -0.519 & 0.057 & 0.003 \\ 
8276.60 & H\,{\sc i} (P32) & 8276.31 & -0.520 & 0.070 & 0.004 \\ 
8281.49 & H\,{\sc i} (P31) & 8281.12 & -0.520 & 0.072 & 0.004 \\ 
8286.71 & H\,{\sc i} (P30) & 8286.43 & -0.521 & 0.111 & 0.005 \\ 
8292.67 & H\,{\sc i} (P29) & 8292.31 & -0.521 & 0.096 & 0.005 \\ 
8299.20 & H\,{\sc i} (P28) & 8298.83 & -0.522 & 0.118 & 0.005 \\ 
8306.48 & H\,{\sc i} (P27) & 8306.11 & -0.523 & 0.119 & 0.005 \\ 
8314.61 & H\,{\sc i} (P26) & 8314.26 & -0.524 & 0.128 & 0.006 \\ 
8323.76 & H\,{\sc i} (P25) & 8323.42 & -0.525 & 0.143 & 0.006 \\ 
8334.11 & H\,{\sc i} (P24) & 8333.78 & -0.526 & 0.152 & 0.007 \\ 
8342.74 & C\,{\sc iii}       & 8342.20 & -0.527 & 0.020 & 0.002 \\ 
8345.85 & H\,{\sc i} (P23) & 8345.55 & -0.527 & 0.188 & 0.008 \\ 
8359.31 & H\,{\sc i} (P22) & 8359.00 & -0.529 & 0.220 & 0.009 \\ 
8362.07 & He\,{\sc i}        & 8361.73 & -0.529 & 0.082 & 0.004 \\ 
8374.81 & H\,{\sc i} (P21) & 8374.48 & -0.531 & 0.219 & 0.009 \\ 
8392.73 & H\,{\sc i} (P20) & 8392.40 & -0.533 & 0.252 & 0.011 \\ 
8397.80 & He\,{\sc i}        & 8397.42 & -0.533 & 0.010 & 0.001 \\ 
8413.65 & H\,{\sc i} (P19) & 8413.32 & -0.535 & 0.288 & 0.012 \\ 
8434.02 & [Cl\,{\sc iii}]    & 8434.00 & -0.537 & 0.008 & 0.001 \\ 
8438.29 & H\,{\sc i} (P18) & 8437.95 & -0.537 & 0.326 & 0.014 \\ 
8444.79 & He\,{\sc i}        & 8444.55 & -0.538 & 0.025 & 0.003 \\ 
8446.82 & O \,{\sc i}         & 8446.48 & -0.538 & 0.468 & 0.020 \\ 
8451.52 & He\,{\sc i}        & 8451.17 & -0.539 & 0.013 & 0.003 \\ 
8467.58 & H\,{\sc i} (P17) & 8467.25 & -0.541 & 0.378 & 0.016 \\ 
8486.63 & He\,{\sc i}        & 8480.79 & -0.543 & 0.017 & 0.002 \\ 
8500.45 & [Cl\,{\sc iii}]    & 8500.20 & -0.544 & 0.010 & 0.004 \\ 
8502.82 & H\,{\sc i} (P16) & 8502.48 & -0.544 & 0.450 & 0.020 \\ 
8545.71 & H\,{\sc i} (P15) & 8545.38 & -0.549 & 0.508 & 0.023 \\ 
8579.06 & [Cl\,{\sc ii}]     & 8578.69 & -0.552 & 0.304 & 0.014 \\ 
8582.21 & He\,{\sc i}        & 8581.88 & -0.552 & 0.024 & 0.003 \\ 
8598.73 & H\,{\sc i} (P14) & 8598.39 & -0.554 & 0.636 & 0.028 \\ 
8617.23 & [Fe\,{\sc ii}]     & 8616.95 & -0.556 & 0.042 & 0.002 \\ 
8648.62 & He\,{\sc i}        & 8648.26 & -0.559 & 0.030 & 0.002 \\ 
8665.37 & H\,{\sc i} (P13) & 8665.02 & -0.560 & 0.829 & 0.037 \\ 
8680.62 & N\,{\sc i}         & 8680.28 & -0.562 & 0.017 & 0.001 \\ 
8683.85 & N\,{\sc i}         & 8683.40 & -0.562 & 0.017 & 0.001 \\ 
8703.64 & N\,{\sc i}         & 8703.25 & -0.564 & 0.012 & 0.001 \\ 
8727.55 & [C\,{\sc i}]       & 8727.12 & -0.566 & 0.014 & 0.002 \\ 
8733.82 & He\,{\sc i}        & 8733.44 & -0.567 & 0.037 & 0.003 \\ 
8736.36 & He\,{\sc i}        & 8736.04 & -0.567 & 0.012 & 0.002 \\ 
8750.82 & H\,{\sc i} (P12) & 8750.47 & -0.568 & 1.042 & 0.047 \\ 
8777.05 & He\,{\sc i}        & 8776.83 & -0.571 & 0.053 & 0.003 \\ 
8845.72 & He\,{\sc i}        & 8845.39 & -0.576 & 0.055 & 0.004 \\ 
8892.21 & [Fe\,{\sc ii}]     & 8891.91 & -0.580 & 0.016 & 0.002 \\ 
8997.34 & He\,{\sc i}        & 8997.00 & -0.588 & 0.055 & 0.004 \\ 
9000.05 & He\,{\sc i}        & 8999.75 & -0.589 & 0.022 & 0.002 \\ 
9015.29 & H\,{\sc i} (P10) & 9014.91 & -0.590 & 1.624 & 0.077 \\ 
9052.33 & [Fe\,{\sc ii}]     & 9051.95 & -0.592 & 0.011 & 0.002 \\ 
9063.71 & He\,{\sc i}        & 9063.32 & -0.593 & 0.060 & 0.004 \\ 
9069.29 & [S\,{\sc iii}]     & 9068.60 & -0.594 & 12.577 & 0.596 \\ 
9123.99 & [Cl\,{\sc ii}]     & 9123.60 & -0.598 & 0.103 & 0.005 \\ 
9210.69 & He\,{\sc i}        & 9210.34 & -0.604 & 0.099 & 0.006 \\ 
   \enddata
 \end{deluxetable}
 
\clearpage

  \begin{table}
   \renewcommand{\thetable}{A\arabic{table}}
  \caption{The identified lines in the 
 \emph{Spitzer}/IRS spectrum. The flux density
 is scaled-up to match with the \emph{AKARI}/IRC 9.0\,{\micron}
 band flux density. Then, the fluxes of these identified atomic lines 
are normalized with respect to the {\hb} flux of the entire nebula 
\mbox{9.83(--12) $\pm$ 7.33(--13)} erg\,s$^{-1}$\,cm$^{-2}$.
 See text in details. \label{T:spitzer}}
\renewcommand{\arraystretch}{0.85}
 \footnotesize
 \centering
\begin{tabularx}{0.7\textwidth}{@{}D{.}{.}{-1}CD{p}{ \pm }{-1}@{}}
 \hline\hline
\multicolumn{1}{c}{$\lambda_{\rm lab.}$} &Line&\multicolumn{1}{c}{$I$($\lambda$)} \\
\multicolumn{1}{c}{($\micron$)} &&\multicolumn{1}{c}{($I$({\hb}) = 100)}       \\
\hline
5.92 & H\,{\sc i} & 0.740~p~0.042 \\ 
6.99 & $[$Ar\,{\sc ii}$]$ & 7.435~p~0.253 \\ 
7.48 & H\,{\sc i} & 6.186~p~0.223 \\
9.01 & $[$Ar\,{\sc iii}$]$ & 12.993~p~0.471\\ 
10.51 & $[$S\,{\sc iv}$]$ & 7.758~p~0.266 \\  
11.31 & H\,{\sc i}& 0.277~p~0.068 \\
12.37 & H\,{\sc i}   & 1.043~p~0.034 \\ 
12.81 & $[$Ne\,{\sc ii}$]$ & 46.711~p~1.553 \\
14.37 & $[$Cl\,{\sc ii}$]$? & 0.323~p~0.040 \\  
15.56 & $[$Ne\,{\sc iii}$]$ & 96.699~p~3.188 \\   
17.62 & He\,{\sc i}? & 1.274~p~0.094 \\  
18.71 & $[$S\,{\sc iii}$]$ & 13.864~p~0.465 \\ 
33.48 & $[$S\,{\sc iii}$]$ & 3.517~p~0.272 \\  
36.02 & $[$Ne\,{\sc iii}$]$ &6.180~p~0.323 \\
 \hline
\end{tabularx}
  \end{table}

  \begin{table*}[t!]
   \renewcommand{\thetable}{A\arabic{table}}
 \caption{Comparison of the overlapped line fluxes between our 2006 
  and \citet{Arkhipova:2013aa}'s 2011 observations. $F$($\lambda$) and
  $I$($\lambda$) are normalized to the $F$({\hb}) and $I$({\hb}), where
  are 100, respectively.
  \label{T:comp_obs}}
 \centering
 \footnotesize
 \renewcommand{\arraystretch}{0.95}
\begin{tabularx}{0.7\textwidth}{@{}CCD{p}{ \pm }{-1}D{p}{ \pm }{-1}D{p}{ \pm }{-1}D{p}{ \pm }{-1}@{}}
 \hline\hline
 $\lambda_{\rm lab.}$ ({\AA})
 &Line&\multicolumn{1}{c}{$F$($\lambda$) in 2006}&
 \multicolumn{1}{c}{$F$($\lambda$) in 2011}
 &\multicolumn{1}{c}{$I$($\lambda$) in 2006}&
 \multicolumn{1}{c}{$I$($\lambda$) in 2011}\\
 \hline
3726/29 & {\oii} &  130.595~p~0.345&126.8~p~7.2&138.857~p~2.801 & 163.0~p~9.5 \\
4101.73 & B6 &  20.596~p~0.031 &18.4~p~2.5&21.516~p~0.395 & 21.6~p~2.9       \\
4340.46 & B5 & 44.700~p~0.067 &45.8~p~2.2&46.052~p~0.579 & 50.9~p~2.5        \\
4363.21 & {\oiii} & 2.392~p~0.004 &4.1~p~1.6&2.461~p~0.030 & 4.6~p~1.8     \\
4958.91 & {\oiii} & 146.232~p~0.237 &150.8~p~3.8&145.519~p~0.380 & 148.0~p~3.8  \\
5754.64 & {\nii} & 2.671~p~0.006 &3.9~p~1.1& 2.578~p~0.038 & 3.3~p~0.9       \\
5875.60 & {\hei} & 15.244~p~0.115 & 18.1~p~0.9&14.666~p~0.262 & 15.2~p~0.7     \\
6300.30 & {\oi} & 16.956~p~0.021 &22.5~p~0.9&16.129~p~0.339 & 17.7~p~0.7      \\
6312.10 & {\siii} &  1.084~p~0.003 &1.6~p~0.5&1.031~p~0.022 & 1.2 ~p~0.4     \\
6363.78 & {\oi} & 5.365~p~0.09 &8.3~p~0.7&5.095~p~0.111 & 6.5~p~0.6         \\
6548.04 & {\nii} & 43.327~p~0.054 &60.9~p~1.6&40.956~p~0.967 & 46.5~p~1.3     \\
6583.46 & {\nii} & 128.601~p~0.168 &190.2~p~5.1& 121.454~p~2.915 & 144.6~p~4.2 \\
6678.15 & {\hei} &  4.215~p~0.007 &5.6~p~0.8&3.972~p~0.099 & 4.2~p~0.6       \\
6716.44 & {\sii} & 6.444~p~0.010 &11.3~p~0.9&6.066~p~0.154 & 8.5~p~0.6       \\
6730.81 & {\sii} &  13.255~p~0.021 &21.6~p~1.0&12.471~p~0.319 & 16.1~p~0.8    \\
7065.18 & {\hei} & 8.410~p~0.101 &10.3~p~1.1&7.846~p~0.247 & 7.4~p~0.8       \\
7135.80 & {\ariii} & 11.562~p~0.018 &17.3~p~0.8&10.767~p~0.322 & 12.3~p~0.6   \\
\hline
\end{tabularx}
  \end{table*}

   \begin{table}
    \renewcommand{\thetable}{A\arabic{table}}
\caption{Adopting {\Ne} and {\te} for the ionic abundance
  derivations. \label{T:nete}}
  \footnotesize
\renewcommand{\arraystretch}{0.95}
 \tablecolumns{6}
\centering
\begin{tabularx}{0.7\textwidth}{@{}LCC@{}}
 \hline\hline
 Ion &{\Ne} (cm$^{-3}$)&{\te} (K)\\
 \hline
N$^{0}$, O$^{0}$ & 1390 & 8470 \\
S$^{+}$ & 5710 & 9280 \\
C$^{2+}$(RL), O$^{2+}$(RL), N$^{2+}$(RL)& 10\,000 & 8090 \\   
He$^{+}$ & 10\,000 & 8160 \\  
 N$^{+}$, O$^{+}$, Cl$^{+}$, Fe$^{2+}$, Ar$^{+}$ & 17\,520 & 9280 \\
S$^{2+}$ & 21\,990 & 8160 \\
 Ar$^{2+}$ & 21\,990 & 8540 \\
Ne$^{2+}$ & 22\,720 & 8560 \\ 
O$^{2+}$(CEL), S$^{3+}$, Ar$^{3+}$ & 22\,750 & 9420 \\ 
Ne$^{+}$ & 22\,980 & 8540 \\ 
Cl$^{2+}$ & 23\,970 & 7490 \\

 \hline
\end{tabularx}
   \end{table}

 \begin{table*}
 \renewcommand{\thetable}{A\arabic{table}}
 \caption{Twice expansion velocities of Hen3-1357. \label{T:exp}}
 \centering
  \footnotesize
\renewcommand{\arraystretch}{0.85}
 \tablewidth{\columnwidth}
\begin{tabularx}{0.7\textwidth}{@{}CCCD{.}{.}{-1}C@{}}
 \hline\hline
 Ion &No. of &Emiss.  &\multicolumn{1}{c}{IP}&2$V_{\rm exp}$\\
     &sample &type     &\multicolumn{1}{c}{(eV)}&({\kms})\\
 \hline
H\,{\sc i} & 39 & RL & 13.60 & 14.78 $\pm$ 0.45 \\ 
O\,{\sc i} & 5 & RL & 13.62 & 21.42 $\pm$ 1.13 \\ 
N\,{\sc i} & 3 & RL & 14.53 & 18.44 $\pm$ 1.42 \\ 
C\,{\sc ii} & 2 & RL & 24.38 & 14.41 $\pm$ 1.51 \\ 
{\hei} & 27 & RL & 24.59 & 16.28 $\pm$ 0.28 \\ 
N\,{\sc ii} & 3 & RL & 29.60 & 14.19 $\pm$ 1.78 \\ 
O\,{\sc ii} & 16 & RL & 35.12 & 16.43 $\pm$ 1.32 \\ 
{\ci} & 1 & CEL & 0.00 & 18.06 $\pm$ 1.84 \\ 
{\Ni} & 2 & CEL & 0.00 & 17.52 $\pm$ 0.20 \\
{\oi} & 3 & CEL & 0.00 & 14.86 $\pm$ 0.08 \\
$[$Fe\,{\sc ii}$]$ & 3 & CEL & 7.87 & 15.25 $\pm$ 1.48 \\ 
{\sii} & 4 & CEL & 10.36 & 14.21 $\pm$ 0.03\\
{\clii} & 2 & CEL & 12.97 & 16.94 $\pm$ 0.23\\
 {\oii} & 6 & CEL & 13.62 & 13.73 $\pm$ 0.05 \\
 {\nii} & 4 & CEL & 14.53 & 14.37 $\pm$ 0.64 \\
 {\feii} & 8 & CEL & 16.18 & 17.97 $\pm$ 1.49 \\
 {\siii} & 3 & CEL & 23.33 & 13.53 $\pm$ 0.03 \\
 {\cliii} & 3 & CEL & 23.81 & 14.17 $\pm$ 0.92 \\
{\ariii} & 3 & CEL & 27.63 & 13.09 $\pm$ 0.23 \\ 
{\oiii} & 3 & CEL & 35.12 & 14.19 $\pm$ 0.26 \\ 
 {\ariv} & 2 & CEL & 40.74 & 16.72 $\pm$ 0.91 \\
 {\neiii} & 2 & CEL & 40.96 & 14.76 $\pm$ 0.04 \\ 
\hline
\end{tabularx}
 \end{table*}

 \begin{table*}
  \renewcommand{\thetable}{A\arabic{table}}
 \caption{
 The comparison between the observed and model
  predicted line fluxes, band fluxes, and band flux
  densities. $\Delta$\,$\lambda$ indicates the bandwidth of each band. The
  predicted $F_{\nu}$ at IRAC-1/2/3/4 bands is 8.56, 1.01(+1), 8.62, and
  3.70(+1) mJy, respectively. The $F_{\nu}$ at IRC-S9W/L18W bands is 1.03(+2) and 2.66(+3) mJy, respectively.
  The $F_{\nu}$ at FIS-N90/WS/WL is 2.25(+3), 1.46(+3), 3.03(+2) mJy, respectively. 
 \label{T:model2}}
 \renewcommand{\arraystretch}{0.85}
 \centering
\footnotesize
\begin{tabularx}{0.7\textwidth}{@{}RLRR@{}}  
 \hline\hline
\multicolumn{1}{c}{$\lambda_{\rm lab.}$}&Ion&$I$(obs)&$I$(model)\\
 &        &($I$({\hb})=100)&($I$({\hb})=100)\\
      \hline
3697.2\,{\AA}   &   H\,{\sc i} (B17)   & 1.923 & 1.146 \\ 
3703.9\,{\AA}   &   H\,{\sc i} (B16)  & 2.028 & 1.347 \\ 
3712.0\,{\AA}   &   H\,{\sc i} (B15)  & 2.384 & 1.606 \\ 
3721.9\,{\AA}   &   H\,{\sc i} (B14)  & 4.156 & 1.949 \\ 
3726.0\,{\AA}   &   {\oii}   & 101.210 & 133.154 \\ 
3728.8\,{\AA}   &   {\oii}   & 37.647 & 54.376 \\ 
3734.4\,{\AA}   &   H\,{\sc i} (B13)  & 3.026 & 3.039 \\ 
3750.2\,{\AA}   &   H\,{\sc i} (B12)  & 4.220 & 2.408 \\ 
3770.6\,{\AA}   &   H\,{\sc i} (B11)  & 4.353 & 3.930 \\ 
3797.9\,{\AA}   &   H\,{\sc i} (B10)  & 5.330 & 5.228 \\ 
3819.6\,{\AA}   &   {\hei}   & 1.142 & 1.228 \\ 
3835.4\,{\AA}   &   H\,{\sc i} (B9)  & 7.597 & 7.192 \\ 
3867.5\,{\AA}   &   {\hei}   & 0.147 & 0.107 \\ 
3869.1\,{\AA}   &   {\neiii}   & 40.325 & 35.532 \\ 
3889.1\,{\AA}   &   H\,{\sc i} (B8)  & 15.997 & 10.316 \\ 
3964.7\,{\AA}   &   {\hei}   & 0.540 & 1.055 \\ 
3967.8\,{\AA}   &   {\neiii}   & 10.000 & 10.709 \\ 
3970.1\,{\AA}   &   H\,{\sc i} (B7)  & 16.026 & 15.611 \\ 
4026.2\,{\AA}   &   {\hei}   & 1.532 & 2.229 \\ 
4068.6\,{\AA}   &   {\sii}   & 4.471 & 4.474 \\ 
4075.0\,{\AA}   &   O\,{\sc ii}   & 0.197 & 0.134 \\ 
4076.4\,{\AA}   &   {\sii}   & 1.508 & 1.450 \\ 
4101.7\,{\AA}   &   H\,{\sc i} (B6, H$\delta$)  & 21.516 & 25.961 \\ 
4120.8\,{\AA}   &   {\hei}   & 0.177 & 0.209 \\ 
4143.8\,{\AA}   &   {\hei}   & 0.229 & 0.350 \\ 
4267.2\,{\AA}   &   C\,{\sc ii}   & 0.103 & 0.126 \\ 
4340.5\,{\AA}   &   H\,{\sc i} (B5, H$\gamma$)  & 46.052 & 46.863 \\ 
4363.2\,{\AA}   &   {\oiii}   & 2.461 & 2.190 \\ 
4387.9\,{\AA}   &   {\hei}   & 0.439 & 0.596 \\ 
4437.6\,{\AA}   &   {\hei}   & 0.066 & 0.084 \\ 
4471.5\,{\AA}   &   {\hei}   & 4.628 & 4.853 \\ 
4651.0\,{\AA}   &   O\,{\sc ii}   & 0.315 & 0.138 \\ 
4658.1\,{\AA}   &   {\feiii}   & 0.111 & 0.222 \\ 
4701.5\,{\AA}   &   {\feiii}   & 0.046 & 0.054 \\ 
4711.4\,{\AA}   &   {\ariv}   & 0.034 & 0.021 \\ 
  4713.2\,{\AA}   &   {\hei}   & 0.668 & 0.662 \\ 
  4733.9\,{\AA}   &   {\feiii}   & 0.026 & 0.022 \\ 
  4740.2\,{\AA}   &   {\ariv}   & 0.070 & 0.034 \\ 
  4754.7\,{\AA}   &   {\feiii}   & 0.029 & 0.025 \\ 
  4881.0\,{\AA}   &   {\feiii}   & 0.045 & 0.065 \\ 
  4921.9\,{\AA}   &   {\hei}   & 1.214 & 1.293 \\ 
  4931.2\,{\AA}   &   {\oiii}   & 0.055 & 0.054 \\ 
  4958.9\,{\AA}   &   {\oiii}   & 145.519 & 131.766 \\ 
 4987.2\,{\AA}   &   {\feiii}   & 0.016 & 0.011 \\
 5015.7\,{\AA}   &   {\hei}   & 2.161 & 2.655 \\ 
  5047.7\,{\AA}   &   {\hei}   & 0.168 & 0.210 \\ 
  5191.8\,{\AA}   &   {\ariii}   & 0.064 & 0.137 \\ 
  5197.9\,{\AA}   &   {\Ni}   & 0.363 & 0.248 \\ 
  5200.3\,{\AA}   &   {\Ni}   & 0.228 & 0.152 \\ 
  5270.4\,{\AA}   &   {\feiii}   & 0.059 & 0.081 \\ 
  5517.7\,{\AA}   &   {\cliii}   & 0.117 & 0.191 \\ 
  5537.9\,{\AA}   &   {\cliii}   & 0.240 & 0.325 \\ 
  5577.3\,{\AA}   &   {\oi}   & 0.219 & 0.142 \\ 
  5679.6\,{\AA}   &   N\,{\sc ii}   & 0.017 & 0.018 \\ 
  5754.6\,{\AA}   &   {\nii}   & 2.578 & 2.946 \\ 
  5875.6\,{\AA}   &   {\hei}   & 14.666 & 15.010 \\ 
  6300.3\,{\AA}   &   {\oi}   & 16.129 & 7.661 \\ 
  6312.1\,{\AA}   &   {\siii}   & 1.031 & 1.850 \\ 
  6363.8\,{\AA}   &   {\oi}   & 5.095 & 2.443 \\ 
  6548.0\,{\AA}   &   {\nii}   & 40.956 & 44.231 \\ 
  6583.5\,{\AA}   &   {\nii}   & 121.454 & 130.527 \\ 
  6678.2\,{\AA}   &   {\hei}   & 3.972 & 3.982 \\ 
 \hline
\end{tabularx}
 \end{table*}  
\setcounter{table}{5}
  \begin{table*}
   \renewcommand{\thetable}{A\arabic{table}}
  \caption{(Continued)}
\renewcommand{\arraystretch}{0.85}
\centering
\footnotesize
\begin{tabularx}{0.7\textwidth}{@{}RLRR@{}}  
\hline
\multicolumn{1}{c}{$\lambda_{\rm lab.}$}&Ion&$I$(obs)&$I$(model)\\
 &        &($I$({\hb})=100)&($I$({\hb})=100)\\
 \hline
  6716.4\,{\AA}   &   {\sii}   & 6.066 & 3.861 \\ 
  6730.8\,{\AA}   &   {\sii}   & 12.471 & 7.635 \\ 
  7065.2\,{\AA}   &   {\hei}   & 7.846 & 7.238 \\ 
  7135.8\,{\AA}   &   {\ariii}   & 10.767 & 19.209 \\ 
  7281.4\,{\AA}   &   {\hei}   & 0.726 & 0.880 \\ 
  7319/20\,{\AA}   &   {\oii}   & 17.485 & 17.826 \\ 
  7329/30\,{\AA}   &   {\oii}   & 14.463 & 14.279 \\ 
  7751.1\,{\AA}   &   {\ariii}   & 2.630 & 4.635 \\ 
  8333.8\,{\AA}   &   H\,{\sc i} (P24)   & 0.152 & 0.158 \\ 
  8345.6\,{\AA}   &   H\,{\sc i} (P23)  & 0.188 & 0.175 \\ 
  8359.0\,{\AA}   &   H\,{\sc i} (P22)  & 0.220 & 0.195 \\ 
  8361.7\,{\AA}   &   {\hei}   & 0.082 & 0.098 \\ 
  8374.5\,{\AA}   &   H\,{\sc i} (P21)  & 0.219 & 0.219 \\ 
  8392.4\,{\AA}   &   H\,{\sc i} (P20)  & 0.252 & 0.248 \\ 
  8413.3\,{\AA}   &   H\,{\sc i} (P19)  & 0.288 & 0.284 \\ 
  8434.0\,{\AA}   &   {\cliii}   & 0.008 & 0.013 \\ 
  8438.0\,{\AA}   &   H\,{\sc i} (P18)  & 0.326 & 0.328 \\ 
  8467.3\,{\AA}   &   H\,{\sc i} (P17)  & 0.378 & 0.383 \\ 
  8500.2\,{\AA}   &   {\cliii}   & 0.010 & 0.015 \\ 
  8502.5\,{\AA}   &   H\,{\sc i} (P16)  & 0.450 & 0.454 \\ 
  8545.4\,{\AA}   &   H\,{\sc i} (P15)  & 0.508 & 0.545 \\ 
  8578.7\,{\AA}   &   {\clii}   & 0.304 & 0.200 \\ 
  8598.4\,{\AA}   &   H\,{\sc i} (P14)  & 0.636 & 0.665 \\ 
  8617.0\,{\AA}   &   {\feii}       & 0.042 & 0.031 \\ 
  8665.0\,{\AA}   &   H\,{\sc i} (P13)  & 0.829 & 0.827 \\ 
  8727.1\,{\AA}   &   {\ci}   & 0.014 & 0.019 \\ 
  8750.5\,{\AA}   &   H\,{\sc i} (P12)  & 1.042 & 1.049 \\ 
  8891.9\,{\AA}   &   {\feii}   & 0.016 & 0.010 \\ 
  9014.9\,{\AA}   &   H\,{\sc i} (P10)  & 1.624 & 1.815 \\ 
9052.0\,{\AA}   &   {\feii}   & 0.011 & 0.007 \\ 
  9068.6\,{\AA}   &   {\siii}   & 12.577 & 27.975 \\ 
  9123.6\,{\AA}   &   {\clii}   & 0.103 & 0.052 \\ 
  5.92\,{\micron}   &   H\,{\sc i}   & 0.740 & 0.464 \\ 
  6.99\,{\micron}   &   {\arii}   & 7.435 & 1.015 \\ 
  9.01\,{\micron}   &   {\ariii}   & 12.993 & 17.582 \\ 
   10.51\,{\micron}   &   {\siv}   & 7.758 & 3.728 \\ 
  11.31\,{\micron}   &   H\,{\sc i}   & 0.277 & 0.308 \\ 
   12.37\,{\micron}   &   H\,{\sc i}   & 1.043 & 0.971 \\ 
  12.81\,{\micron}   &   {\neii}   & 46.711 & 85.274 \\ 
  15.56\,{\micron}   &   {\neiii}   & 96.699 & 54.536 \\ 
  18.71\,{\micron}   &   {\siii}   & 13.864 & 20.088 \\ 
  33.48\,{\micron}   &   {\siii}   & 3.517 & 4.012 \\ 
  36.02\,{\micron}   &   {\neiii}   & 6.180 & 3.746 \\   
\hline 
\multicolumn{1}{c}{$\lambda_{c}$($\Delta$\,$\lambda$)}&Band&$I$(obs)&$I$(model)\\
                                                       &    &($I$({\hb})=100)&($I$({\hb})=100)\\
\hline
3.56(0.68)\,{\micron}   &   IRAC-1   & 17.926 & 14.099 \\ 
4.51(0.86)\,{\micron}   &   IRAC-2   & 20.919 & 13.128 \\ 
5.74(1.26)\,{\micron}   &   IRAC-3   & 12.618 & 10.021 \\ 
7.93(2.53)\,{\micron}   &   IRAC-4   & 48.754 & 45.385 \\ 
9.22(4.10)\,{\micron}   &   IRC-S9W   & 130.819 & 152.073 \\ 
19.81(9.97)\,{\micron}    &   IRC-L18W   & 1990.469 & 2062.191 \\ 
65.0(20.17)\,{\micron}    &   FIS-N60   & 327.889 & 327.959 \\ 
90.0(39.90)\,{\micron}    &   FIS-WS   & 283.292 & 219.030 \\ 
140.0(54.74)\,{\micron}   &   FIS-WL   & 32.158 & 25.831 \\ 
\hline
 \multicolumn{1}{c}{$\lambda_{c}$($\Delta$\,$\lambda$)}&Band&$I$(obs)&$I$(model)\\
                                                       &    &($I$({\hb})=100)&($I$({\hb})=100)\\
\hline
  8.20(0.30)\,{\micron}   &   IRS-1   & 1.613 & 2.279 \\ 
  9.55(0.10)\,{\micron}   &   IRS-2   & 3.008 & 4.012 \\ 
  10.95(0.50)\,{\micron}   &   IRS-3   & 21.065 & 29.324 \\ 
  14.95(0.50)\,{\micron}   &   IRS-4   & 29.190 & 41.669 \\ 
  16.70(0.60)\,{\micron}   &   IRS-5   & 98.078 & 95.574 \\ 
  18.10(0.60)\,{\micron}   &   IRS-6   & 128.928 & 132.281 \\ 
  22.50(1.00)\,{\micron}   &   IRS-7   & 215.444 & 217.328 \\ 
  29.50(1.00)\,{\micron}   &   IRS-8   & 127.541 & 150.879 \\ 
 \hline
\multicolumn{1}{c}{$\lambda_{c}$} & &$F_{\nu}$(obs)&$F_{\nu}$(model)\\
                                  & & (Jy)&(Jy)\\             
 \hline
 9.60\,{\micron}   &      & 0.097 & 0.110 \\ 
  17.60\,{\micron}   &      & 2.209 & 1.898 \\ 
  25.00\,{\micron}   &      & 3.718 & 3.746 \\ 
  27.00\,{\micron}   &      & 3.695 & 3.943 \\ 
  29.00\,{\micron}   &      & 3.790 & 4.060 \\ 
\hline
\end{tabularx}
  \end{table*}

\end{document}